\documentclass[a4paper,12pt]{article}
\usepackage{amsfonts,amsthm,amsmath,amssymb,graphicx,hyperref}

\newcommand{\tr}{\operatorname{tr}}

\def\Tr{{\rm Tr\, }}

\def\b{\beta}

\def\g{\gamma}

\def\s{\sigma}

\def\d{\partial}

\def\t{\tau }

\def\hg{ \hat{g}}

\newcommand{\cF}{\mathcal F}

\newcommand{\cM}{\mathcal M}

\newcommand{\cO}{\mathcal O}

\newcommand{\bX}{{\bf X }}

\newcommand{\be}{\begin{equation}}
\newcommand{\bea}{\begin{eqnarray}}

\newcommand{\ee}{\end{equation}}
\newcommand{\eea}{\end{eqnarray}}

\newcommand\bdlt{ \boldsymbol{ \delta }} 
\newcommand\bfe{ {\bf e}  }

\def\bT{ { \bf{T}}  }
\def\g{\gamma}
\def\s{ \sigma}

\def\mQ{ { \mathbb{Q} } }
\def\bmQ{ \bar { \mathbb{Q}}}
\def\mbC{ \mathbb{C}}
\def\mP{ \mathbb{P}}

\begin{document}

\makeatletter
\@addtoreset{equation}{section}
\makeatother
\renewcommand{\theequation}{\thesection.\arabic{equation}}

\rightline{QMUL-PH-09-21}
\rightline{WITS-CTP-051}
   \vspace{1.8truecm}

\vspace{15pt}


\centerline{\LARGE \bf  From Matrix Models and   quantum fields }
\centerline{ \LARGE\bf  to Hurwitz space and the absolute Galois group } \vspace{1truecm}
\thispagestyle{empty} \centerline{
    {\large \bf Robert de Mello Koch
${}^{a,} $\footnote{ {\tt robert@neo.phys.wits.ac.z}}}
   {\large \bf and Sanjaye Ramgoolam
               ${}^{b,}$\footnote{ {\tt s.ramgoolam@qmul.ac.uk}} }
                                                       }

\vspace{.4cm}
\centerline{{\it ${}^a$ National Institute for Theoretical Physics ,}}
\centerline{{\it Department of Physics and Centre for Theoretical Physics }}
\centerline{{\it University of Witwatersrand, Wits, 2050, } }
\centerline{{\it South Africa } }

\vspace{.4cm}
\centerline{{\it ${}^b$ Centre for Research in String Theory, Department of Physics},}
\centerline{{ \it Queen Mary University of London},} \centerline{{\it
    Mile End Road, London E1 4NS, UK}}

\vspace{1.4truecm}

\thispagestyle{empty}

\centerline{\bf ABSTRACT}

\vspace{.4truecm}

\noindent

We show that correlators of the hermitian one-Matrix model with a general potential 
can be mapped to the counting  of certain triples of permutations and
hence to  counting of holomorphic maps from world-sheet to 
 sphere target with  three branch points on the target. 
This allows the use of old matrix model results to
derive new explicit  formulae for a class of   Hurwitz numbers.
Holomorphic maps  with three branch points
are related,  by Belyi's theorem,  to curves and maps 
defined over algebraic numbers $\bmQ$. This
shows that the  string theory dual of the one-matrix model at 
generic couplings has worldsheets defined 
 over the algebraic numbers and a target space $ \mP^1 ( \bmQ )$. 
The absolute  Galois group $ Gal ( \bmQ / \mQ ) $
acts on the Feynman diagrams of the 1-matrix model, which
are related to Grothendieck's Dessins d'Enfants. 
Correlators of  multi-matrix models are mapped 
to the  counting of triples of permutations
subject to equivalences defined by subgroups of the permutation groups.
This is related to colorings of the edges of the Grothendieck Dessins.
The colored-edge Dessins are  useful as a tool for 
describing some known invariants of the $ Gal ( \bmQ / \mQ ) $
action on  Grothendieck Dessins and for defining new invariants.

\vspace{.5cm}

\setcounter{page}{0}
\setcounter{tocdepth}{2}

\newpage

\tableofcontents

\setcounter{footnote}{0}

\linespread{1.1}
\parskip 4pt

{}~
{}~

\newpage

\section{ Introduction }

Hermitian matrix models (subsequently labeled old matrix models) 
 were the centre of intense 
research in string theory in the early nineties
\cite{bk,gm,ds}. This lead to connections 
with 2D topological gravity and intersection theory 
on $ \cM_{g,n}$ \cite{topgrav,topgravi}. A physical picture 
in terms of string theory in one physical (Liouville)
 dimension was developed. Some reviews on the subject are 
 \cite{Ginsparg:1993is,Polchinski:1994mb,dgz}.

In the mid-nineties, a string theory of two dimensional 
Yang Mills theory (2dYM) was discovered \cite{gt}. Exact answers for partition 
functions were converted using Schur-Weyl duality, to 
a $1/N$ expansion where the contribution at each order was expressed 
in terms of symmetric group data. The simplest way to exhibit 
the string theory was in the form of Hurwitz spaces 
of holomorphic maps and Euler characters of these spaces were identified
in the large $N$ expansion \cite{cmr}. A review and references can be found 
in \cite{cmrii}.  
Some recent developments include a
new understanding of the coupled expansion of 2dYM 
in terms of holomorphic maps \cite{kram2},  connections 
between Hurwitz spaces and gauge-string duality  
in higher dimensions ($AdS_3/CFT_2$) \cite{prs0905}
and instanton based methods for the large order behavior 
for certain  Hurwitz spaces with simple branch points \cite{msw}.

In this paper we follow the strategy which proved fruitful 
in constructing the string theory dual of 2dYM : 
express the computation of 
correlation functions in hermitian matrix models 
in terms of symmetric group data and interpret the result 
in terms of branched covers using classic mathematical 
results (Riemann existence theorem). The Riemann surfaces 
appearing as  covering spaces are the string worldsheets. 
The target space of the maps define the target space of the string theory.

In deriving the connection between matrix model 
correlators and symmetric group data, we find 
it useful to use diagrammatic tensor space techniques 
which have found various
applications in 2dYM \cite{srwlp} and 
more recently in AdS/CFT \cite{cjr,cr,dss,
bhr, kr,  rob,bcd}. These techniques  have been
used to compute Wilson loops in 2dYM, diagonal bases
for correlators in the half-BPS sector of $N=4$ SYM
and more general sectors.
We review some relevant aspects of the tensor space techniques
in section (\ref{revtst}).

Our  first result  is a
connection between one-point functions of
arbitrary multi-trace operators  of the one-matrix model
and a counting of equivalence classes of
{\bf three permutations}. These equivalence classes
are defined in  (\ref{revrh}). The Riemann existence theorem 
relates the  counting of strings of permutations to 
 Hurwitz counting problems of holomorphic maps from 
one Riemann surface to another, 
with specified branching data.  We  review this in (\ref{revrh}).

The importance of Riemann's existence theorem is that it relates
a discrete symmetric group  counting  problem
to  holomorphic map  counting defined in the continuum.
The philosophy of the old matrix models was
that the Matrix models generate discretized
Riemann surfaces. Different scaling limits
from these discretizations approach different
 string backgrounds. Here we are using the Riemann existence theorem
 to get at a continuum picture for any correlator for generic potential.
Then we are getting the different string backgrounds
from the scaling limits of this continuum problem
(which admits a mathematically equivalent discrete permutation
 interpretation).

The fact that three permutations appear in the counting
problem means that we are counting branched covers
with 3 branch points on the sphere $\mP^1$,
 which can be chosen to be at $0,1,\infty$.
 The covering space is a Riemann surface, equipped with a map to the
target sphere. The inverse image of each branch point
contains one or more points on the covering Riemann surface
 where the derivative of the map vanishes. 
These  are called ramification points. They are each labeled by a positive 
integer. The branch point is associated with a set of
ramification points and thus with a set of positive integers.
 A remarkable theorem in mathematics \cite{Belyi},
the {\bf Belyi theorem},  implies that, for the 
case of three branch points,  the covering
curve and the map are {\bf defined over algebraic numbers}.
These are numbers which solve polynomial equations
with coefficients in $ \mQ $, the rational numbers.
They give rise to a field  $ \bmQ$, which is the  algebraic closure
of $ \mQ$. An important group in number theory, called the
absolute Galois group $ Gal ( \bmQ / \mQ ) $ (which we will 
often call ``the Galois group''),  acts on $ \bmQ$ while 
leaving $ \mQ$ fixed) organizes all the
key properties of the algebraic numbers
 \cite{absgalois}.  A fact related to Belyi's
theorem, highlighted by Grothendieck, is that the Galois group
acts faithfully on the equivalence classes of  triples of permutations.
He described Dessins D'enfants which capture these equivalence classes.
In our case, the triples are coming from Feynman diagrams of
the one-matrix model. And in fact the direct connection between
Feyman diagrams and Dessins is not hard to see. It is explained
in section (\ref{fromFeynDess}) . This means that the Galois group
acts on the Feynman diagrams, and sets of { \it Feynman diagrams
can be assembled into Galois orbits.} Since the 
Feynman diagrams  also correspond to Hurwitz classes 
consisting of a string worldsheet and a holomorphic map to $\mP^1$,
  we may say that {\it the Galois group organizes 
the string worldsheets} contributing to the sum over maps.

 We generalize the connection between
 permutation triples and correlators to
 multi-matrix models. The Dessins d'Enfants
 of Grothendieck are replaced by { \bf colored-edge  Dessins}.
 We define the equivalence classes of the
 colored-edge Dessins in terms of equivalences generated
 by some subgroups of the permutation group.
 The continuum data related to the colored Dessins
 is shown to be richer than just the holomorphic maps
 $f$ related to the Dessins. It is replaced by pairs $ (f,s)$
 where the holomorphic map $f$ is accompanied
 by additional data consisting of sections of  sheaves
on the covering Riemann surface, supported at some of the
 ramification points. These pairs are naturally related to
 sheaves over Hurwitz space.
We describe these results in section (\ref{shcodes}).

An important problem considered at length in the Math literature
is that of finding { \bf Galois invariants}, properties of Dessins 
which are invariant under the Galois action. 
We provide a new construction
of Galois invariants using the colored-edge Dessins of the
multi-Matrix models. These invariants
can be  defined in terms of  lists of multi-matrix operators, 
which can be viewed combinatorially as
 multiple necklaces-with-colored-beads. We also use the 
colored-edge Dessins to describe some known invariants 
from the mathematical literature.

It is interesting that the traditional picture of old Matrix Models 
is that of discretized worldsheet Riemann surfaces 
for generic potential giving rise to continuum Riemann surfaces in 
double scaling limits, while the picture developed using Hurwitz space and 
 Belyi's  theorem implies that we are getting
 another kind of ``discretization'' for generic potentials, namely
replacing curves, maps and target space  defined over  $\mbC$
with curves, maps and target space defined over $ \bmQ $.

Since this paper overlaps with (superficially) disconnected 
areas of string theory, matrix models and 
number theory, we collect some key words and 
facts in the Appendix \ref{glosskey}, which should be useful 
to diverse readers.

\section{  One  Matrix Model and Hurwitz space }\label{onemathurwitz}

\subsection{ Review :  Riemann's existence theorem and Riemann-Hurwitz formula }\label{revrh}

Using local complex coordinates  a holomorphic map
satisfies $ \bar \d_z f = 0$.
The Riemann existence theorem  relates to the counting of holomorphic maps
$f : \Sigma_h \rightarrow  \Sigma_T$ between world-sheet Riemann 
surface $ \Sigma_h $ of genus $h$ and target space Riemann surface $\Sigma_T$. 
In this paper $ \Sigma_T $ will have genus $0$, so it is the sphere or complex 
projective line 
$ \mP^1 $.  
  Two maps $f_1$ and $f_2$ are defined to
be equivalent if  there is a biholomorphic isomorphism  
$ \phi : \Sigma_h  \rightarrow \Sigma_h  $ such that the following diagram is commutative
\bea
&&   ~~~~ \Sigma_h   \quad ~~ \underrightarrow{\phi }    \quad  ~ \Sigma_h   \cr
&&    \quad f_1   \searrow \qquad  \swarrow f_2  \cr
&&        \qquad \quad    ~~    \mP^1 
\eea
In other words, equivalent maps $ f_1 , f_2 $ obey the 
equation $ f_1 = f_2 \circ \phi $ or $ f_2 = f_1 \circ \phi^{-1}$.   
For a generic point on $P$, the inverse image consists of $d$
points, where $d$ is the degree of the map.
For a finite set of points, called { \bf branch points},
there are fewer inverse images. If we consider a small disc
around a branch point, the inverse images will be a number of
discs. Restricting the map to one of these discs, it looks
like $ w = z^{r}$ for some positive integer $r$. If $r=1$ the
inverse image is an ordinary point. For $r>1 $ it is a { \bf ramification
point}. For $r=2$, it is a simple ramification point.  The vector $(r_1, r_2 , \cdots  )$  called the 
{\bf ramification profile} 
determines  branching numbers $ \sum_{i} ( r_i-1 ) $. The sum of
these over all branch points is denoted as $B$.
The degree of the map is equal to $ d = \sum_{ i}  i r_i  $.
The Riemann-Hurwitz theorem states that
\bea
(2h-2)  = d ( 2G -2 ) + B
\eea
Equivalent maps have the same set of branch points.
Given such a holomorphic map (branched cover) with $L$
branch points, we can get a sequence $ \s_1 , \s_2 , \cdots \s_L $
of permutations in $S_d$. They are obtained by labeling the inverse
images of a generic base point, and following the inverse images of
a closed path starting at the base point. For the sphere target,
we have
\bea
 \s_1 \s_2  \cdots \s_L = 1
\eea
which follows from the fact that a path going round all
the branch points is contractible.
Two equivalent holomorphic maps are described by permutations
$ \s_1  , \s_2 , \cdots , \s_L $ and  $ \s_1^{\prime} , \cdots , \s_L^{\prime} $
which are related by
\bea
\s_i = \alpha \s_i^{\prime} \alpha^{-1}
\eea
for some fixed $ \alpha \in S_d $.
This correspondence between sequences of permutations and 
Holomorphic maps is the { \bf Riemann existence theorem}. 
This reduces the counting of maps with fixed branch points
to a combinatoric problem in permutation groups.
Thus a space defined by maps from a smooth space (complex manifold)
can be described by the data of branch point locations
and some discrete data of permutation counting. For fixed
branch points, the continuum problem is entirely reduced
to discrete data. We will next show that, precisely
this kind of permutation counting arises in the computation of
general 1-point functions in the one-matrix model. In fact we always have
$L=3$.

We recap the key points for the case $L=3$.
Holomorphic maps from Riemann surfaces to
sphere with three branch points are determined 
by three permutations $ \s_1 , \s_2 , \s_3$ such that 
\bea\label{3permseq1}  
\s_1 \s_2 \s_3 = 1 
\eea 
Permutations $ ( \s_1^{\prime} ,\s_2^{\prime}  , \s_3^{\prime}  )$
determine the same map iff 
\bea\label{equivcase3} 
 \s_1^{\prime} = \alpha  \s_1 \alpha^{-1}  \cr 
 \s_2^{\prime} = \alpha  \s_2 \alpha^{-1} 
\eea
for some $ \alpha \in S_{d}$. 
Because of (\ref{3permseq1}), the condition  (\ref{equivcase3}) 
suffices to ensure that $\s_3^{\prime} = \alpha  \s_3 \alpha^{-1}$. 
We will refer to $ ( \s_1 , \s_2 , \s_3 ) $ 
as { \bf Hurwitz data}  for a holomorphic map. Equivalence 
classes under the conjugation (\ref{equivcase3}) will  be 
called { \bf Hurwitz classes}.

The counting of triples obeying (\ref{3permseq1}) 
is conveniently written by defining a delta function 
over symmetric groups.  For $ \sigma \in S_d$
 we define 
\bea\label{delsd}  
  \delta_{S_d}  ( \sigma ) &=& 
 1  ~~ \hbox{if }  \sigma =1 , ~ \hbox{the identity permutation} \cr 
                           &=& 0 ~~ \hbox{if } \sigma \ne 1
\eea 
 By linearity this extends to a delta-function 
on the group algebra. So counting triples is given 
by 
\bea 
\sum_{\s_1 , \s_2 , \s_3 \in S_d } \delta_{S_d} ( \s_1 \s_2 \s_3 ) 
\eea

\subsection{ Brief review of diagrams and  tensor space techniques }\label{revtst}

\begin{figure}
\begin{center}
 \resizebox{!}{5cm}{\includegraphics{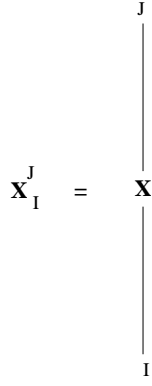}}
\caption{Operator in tensor space and diagram }
 \label{fig:tensdiag1}
\end{center}
\end{figure}

It is useful in computations of correlators in matrix theories,
to think of the matrices $ X , Y , Z $ etc. as operators
in a vector space $V$. Indeed a matrix $X$ is a linear operator
on a vector space $V$. Choosing a basis $|e_i \rangle  $ we
have
\bea
X |e_i > = X_i^j | e_j \rangle
\eea
We can extend this to define an action on $ V^{ \otimes m  }$
as follows
\bea
( X \otimes X \cdots \otimes X )
 |e_{i_1} \otimes e_{i_2} \otimes \cdots e_{i_m}  >
= X_{i_1}^{j_1} \cdots X_{i_m}^{j_m}
|e_{j_1} \otimes e_{j_2} \otimes \cdots e_{j_m}  >
\eea
We can write this more compactly, by writing
$ { \bf X } = ( X \otimes X \cdots \otimes X )$
and the multi-indices  $ I = ( i_1 , i_2 , \cdots , i_n ) $
\bea
{ \bf X } | { \bf e }_{ I } \rangle =  { \bf X }^J_I |e_J >
\eea
By introducing dual vectors, we may also write
\bea
\langle e^J | { \bf X } | e_I \rangle =  { \bf X }^J_I
\eea
Many manipulations are conveniently conducted by
using diagrams. The first step is simply to write the
above operator in tensor space as in the Figure \ref{fig:tensdiag1}.

The strands represent the states (vectors) of $V^{\otimes m }$.

\begin{figure}
\begin{center}
 \resizebox{!}{5cm}{\includegraphics{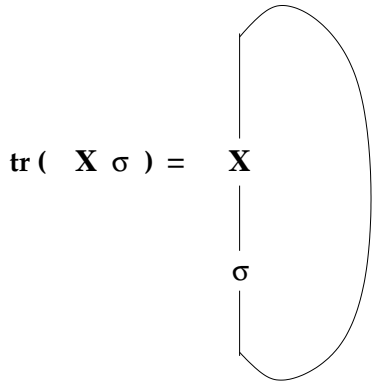}}
\caption{Multitrace operators }
 \label{fig:tensdiag2}
\end{center}
\end{figure}

Different traces and products of traces of $X$, such as
$ ( tr X )^2 tr ( X^2 )$ can be written by composing the
action of ${ \bf X } $ with that of permutations $ \sigma \in S_m $
acting as
\bea
\sigma | e_{i_1} \otimes e_{i_2} \otimes \cdots e_{i_m} \rangle
= |  e_{i_{\s (1) } } \otimes e_{i_{ \s (2)} } \otimes \cdots e_{i_{\s(m)} } \rangle
\eea
For example 
\bea 
tr ( X^2 ) = tr_{ V^{\otimes 2 } }\left(   ( X \otimes X )  (12) \right) = tr_{ V^{\otimes 2 } }\left(  (12)  ( X \otimes X )   \right) 
\eea 
 while
replacing $ (12)$ by the identity permutation gives $( tr X)^2$.
In the diagrammatic representation of tensor space manipulations,
tracing is drawn by joining  strands. Any multi-trace operator
with $m$ copies of $X$ can be obtained from an appropriate permutation
of the strands. This is shown in Figure  \ref{fig:tensdiag2}.
Two permutations which are conjugate to each other give rise 
to the same multi-trace operator. There is a one-one
 correspondence between multi-traces and
conjugacy classes of permutations.

\begin{figure}
\begin{center}
 \resizebox{!}{5cm}{\includegraphics{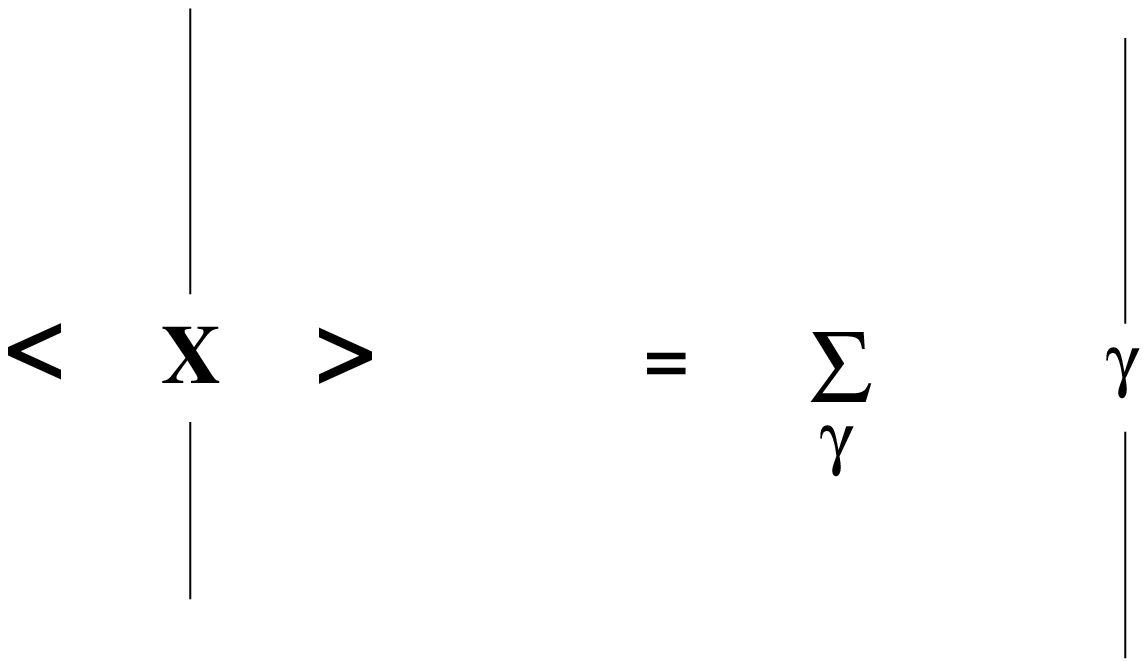}}
\caption{Correlator using Wick's theorem : $ \gamma $ in $ [2^n] $ (see text for further explanation)    }
 \label{fig:tensdiag3}
\end{center}
\end{figure}

In quantum field theory $X$ is a function of space-time
coordinates, and in fact an operator in a Hilbert space,
hence the  terminology ``multi-trace operators''.
Matrix models where $X$ depends on no coordinates at all
are special cases of QFT in zero space-time dimensions.
Most of this paper is indeed focused on that case, but
diagrammatic tensor space techniques are useful more generally.
The key element of free QFT (Gaussian matrix models) we will use
is that observables are correlators of multi-traces and these
can be computed by combining Wick's theorem with the basic formula
\bea\label{basiccorr}
 { \bf \langle }  X^i_j  X^k_l { \bf \rangle } = \delta^i_l \delta^k_j
\eea
Wick's theorem implies that for a correlator
involving a large number of $X$'s we need to
sum
\bea\label{wick}
 { \bf \langle } X^{i_1}_{j_1}  X^{i_2}_{j_2} \cdots   X^{i_{2n}}_{j_{2n} }
 { \bf \rangle }
= \sum_{\gamma \in [2^n] }
   \delta^{i_1}_{ j_{\gamma(1)} }  \delta^{i_2}_{ j_{\gamma(2)}}
   \cdots \delta^{i_{2n} }_{ j_{\gamma(2 n)}}
\eea
The permutation $\gamma $ is being summed over
all elements in  the conjugacy class of
$S_{2n}$ with $n$ cycles of length $2$. The size of the conjugacy class is
$(2n)! \over 2^n n! $ which is the number of ways of choosing
$n$ pairs from $2n$ objects.
 The equation (\ref{wick})  is expressed
diagrammatically in Figure \ref{fig:tensdiag3}.

For computations in Gaussian multi-matrix models, involving
matrices $ X , Y , Z  \cdots $,  the correlator of a pair
of different matrices is zero, and for each pair of like matrices we have
the correlator in (\ref{basiccorr}).
Again, the above remark that multi-traces can be obtained
by composing permutations with an appropriate operator in
tensor space holds true. Multi-traces with $ m_1 , m_2 , m_3  $
copies of $ X , Y , Z  $ are obtained from
$ { \bf X }  \otimes { \bf Y}  \otimes { \bf Z }  $ acting on
$ V^{ \otimes m_1  + m_2 + m_3 } $. While in the case of the
single matrix model,  conjugations  by permutations in $S_{m }$
lead to the same operator, here conjugations by permutations
in the subgroup $ S_{ m_1 } \times S_{m_2} \times S_{m_3} $
lead to the same multi-matrix operator. For a non-zero correlator, 
we need $ m_1 ,m_2,m_3$ to be even, so we write $m_1 = 2n_1, m_2 = 2n_2 , m_3 = 2n_3 $.  Wick's theorem results in
sums over permutations $ \gamma $ in $ [ 2^{2n_1} , 2^{2n_2} , 2^{2n_3} ]  $
of $S_{2n_1} \times  S_{2n_2}  \times S_{2n_3}$.

Some recent papers where diagrammatic tensor space techniques
play an important role include \cite{bhr,kr,rob,bcd,tomloop,Koch:2008cm,krt}.

\subsection{ Gaussian Matrix Model and maps to $\mP^1$ with three branch points   }
\label{OWP-3BPs}

Choose a permutation $ \s \in S_{2n} $ which characterizes a
multi-trace operator with $ 2n $ copies of $ X$. The correlator only
depends on the conjugacy class $ [\s ] $  of $\s $.
Compute the correlator in the Gaussian matrix model which has the two point function
\bea\label{bascor} 
\langle X^{i}_j X^k_l  \rangle  =  \delta^i_l \delta_j^k
\eea
Using this two point function and Wick's theorem (\ref{wick}), we have  
\bea\label{basiccorr2}
&& \langle tr_{ 2n }  ( \sigma \bX ) \rangle
= \sum_{ \g \in [2^n]  }
 tr_{2n} ( \g \sigma   ) \cr
&&= \sum_{ \g \in [2^n]   } N^{ C_{ \g \s   } } \cr
&& =  \sum_{ \tau \in S_{2n} }
 \sum_{ \g \in [2^n]  } N^{ C_{ \tau    } }  \delta_{ S_{2n} } ( \sigma \g \t )
\eea
Here $[2^n]  $ is the conjugacy class with $n$ cycles of length $2$ and $C_{ \g \s } $ is the number of disjoint
cycles in the permutation ${ \g \sigma } $. 

A judicious choice of normalization gives
\bea\label{judicious}
&& { |[ \s] |   \over (2n) ! } N^{ C_{\s} - n  }
   \langle tr_{ 2n }  ( \sigma X ) \rangle \cr
&& = { 1 \over (2n)! }    \sum_{ \tau \in S_{2n} } \sum_{ \s \in [ \s] }
 \sum_{ \g \in [2^n]   }   \delta_{ S_{2n} } ( \sigma \g \t )
N^{ C_{\s} + C_{\tau } - n  }
\eea
In the above $ |[\s]|$ is the size of the conjugacy class $ [\s] $.
$ b ( \s ) = 2n - C_{ \s} $ is the branching number of the permutation,
which only depends on the conjugacy class of the permutation.
$ b( [2^n]  ) $ is the branching number of the conjugacy class
with $n$ cycles of length $2$, which is equal to $ n $.

The above sum counts branched covers of the sphere,
with $3$ branched points,  described by permutations 
$ \s , \g , \t$. The power of $N$ keeps track of the
genus of the worldsheet
\bea\label{countgenus} 
&& (2- 2h  ) = 2n ( 2 - 2G  ) - B \cr
&& = 4n - (  2n - C_\s  ) - ( 2n - C_\tau  ) - ( 2n - C_{ [2^n] }   )\cr 
&& = C_{ \sigma } + C_{ \tau } - n 
\eea
where $G$ is the genus of the target, in this case $0$,
 and $B$ is the total branching number.
Using (\ref{countgenus}) and (\ref{judicious}) we have  
\bea\label{corr3point}
&&  { |[ \s] |   \over (2n) ! } N^{ C_{\s} - n   }
   \langle tr_{ 2n }  ( \sigma X ) \rangle 
 = \sum_{  f ( [\s]  , [2^n ]  ) :\Sigma_h \rightarrow \mP^1   }
 { 1 \over | Aut f | } N^{ 2 - 2h  }
\eea
The sum is over the branched covers with three branch points. 
The first branch point is described by    permutations  $ \sigma $ 
in the conjugacy class $ [\sigma] $ defined 
by the observable $ tr_{2n}  ( \sigma \bf X ) $. 
The second branch point, resulting from Wick contractions, 
 is described by  permutations $ \gamma $ in the conjugacy  $[2^n] $. 
The third branch point can be in any conjugacy class
which arises in the product of  $ \sigma $ and $ \gamma $. 
 Given two maps
\bea
&&  \phi :  \Sigma_h \rightarrow   \Sigma_h \cr
&& f :  \Sigma_h \rightarrow \mP^1
\eea
we say that $ \phi \in Aut ( f) $ if
\bea
f \circ \phi = f
\eea
To obtain the result (\ref{corr3point}) we have used the fact that the sum over $\s$,$\g$ and
$\t$ of $\delta_{ S_{2n} } ( \sigma \g \t )$ with a factor of $ { 1 \over (2n)! } $, is
equal to the sum over maps $f ( \s , \bT ) :\Sigma_h \rightarrow \mP^1$ with each map weighted by
${ 1 \over | Aut f | }$ \cite{gt,cmrii}.

If we sum over $ [\sigma]  $ with the above weights we can  define a
generating function
\bea
&& \cF ( N )  = \sum_n \sum_{[\s]\in [S_{2n}] }  { |[ \s] |   \over (2n) ! } N^{ C_{\s} - n   }
   \langle tr_{ 2n }  ( \sigma X ) \rangle \cr
&& =   \sum_{  f (  [2^n] ) :\Sigma_h \rightarrow P_1   }
 { 1 \over | Aut f | } N^{ 2-2h  }
\eea
Here the sum is over maps, with three branch points,
one with ramification profile  $ [2^n]  $, and the other
two arbitrary, and weighted correctly (as in string theory)
by the worldsheet genus.

\subsection{ Comments on integrality }\label{cominteg}

Considering  (\ref{corr3point})  and (\ref{basiccorr2})
and noting that $ { (2n)! \over | [\sigma ] | } $ is the 
number of permutations $ \alpha $  in $ S_{2n}$ which leave $ \sigma $ 
fixed under conjugations,  i.e $ \alpha \sigma \alpha^{-1} = \sigma $, 
we can write this factor as $ |Aut ( \sigma )| $. 
This indicates that it is the order of the subgroup $ Aut ( \sigma )$ 
of $S_{2n}$ which leave $ \sigma $ fixed under conjugation. 
Hence we can write 
\bea\label{trcor}
\langle tr_{2n}  ( \s X ) \rangle =
\sum_{\s \in [\s]}\sum_{ \g \in [2^n] }  N^{ C_{ \gamma \sigma } }
 {  | Aut ( \sigma ) |  \over | Aut f_{ \sigma , \gamma } |  }
\eea
$f_{ \sigma , \gamma }$ is a Hurwitz class  determined 
by the pair $ \sigma , \gamma $.
We know that the LHS is an integer times a power of $N$
because it is a sum over Wick contractions.
We can also see this directly  from the RHS because
$  Aut f_{ \sigma , \gamma } = Aut (  \s  ) \cap Aut ( \g ) $
which means that it is a subgroup of $  Aut (  \s  )$, hence
by group theory must have an order which
divides  $ | Aut (  \s  ) | $. For the same reason, using 
 $ |Aut ( \g )|  =  2^n n!   $, we have   
$ { 2^n n! \over  | Aut f_{ \sigma , \gamma } | } $ is an integer,
which means
\bea
  { 2^n n! \over |Aut  ( \sigma ) | } \langle tr_{2n}  ( \s X ) \rangle
\eea
is a sum of  positive integers.

The factor $   { |  Aut ( \sigma )| \over | Aut f_{ \sigma , \gamma } |  } $
appearing in (\ref{trcor}) 
 has a nice interpretation. Fix a $ \sigma $ to describe
our operator. Pick a $ \gamma $ which determines a Wick contraction.
The pair $ ( \sigma , \gamma ) $ determines 
a Hurwitz class. As $ \gamma $ runs over
its conjugacy class, we sum over
Wick contractions. How many of these Wick contractions
are in the same equivalence class as $ (\sigma , \gamma )$ ?
We can get other permutations $ \tilde \gamma $ in the conjugacy class of
$ \gamma $ by conjugating with $ h \in S_{2n}$.
Note that  other representatives of the same Hurwitz
class are related to $ \sigma , \gamma $ as
 $ ( h \sigma h^{-1} , h \gamma h^{-1} ) $. As we are summing over $ \gamma $
with $ \sigma $ fixed, we are running over pairs
 $ ( \sigma , h \gamma h^{-1} ) $.
For this to be equivalent to $(  \sigma , \tilde \gamma ) $ we need
$ h $ to fix $ \sigma $, i.e $ h \in Aut (  \sigma )  $.
But if $ h \in Aut (\g ) $ as well, then it does not change
the pair. So the number of Wick contractions which
give contributions from the same Hurwitz class
as the one of $ \sigma , \gamma $ is
$ Aut ( \sigma ) / Aut f_{ \sigma \gamma }  $. Finally, the group that leaves
$\s$ and $\g$ fixed is clearly a subgroup of $Aut (\s\g )$ so that
${  | Aut  ( \sigma \gamma )  | \over | Aut f_{  \s , \g } | }   $ is
also an integer.

\subsection{Perturbed model, 3  Branch points,  and new results on Hurwitz numbers }\label{newhurwitz} 

In this section we will perturb the Gaussian matrix model,
 with a potential of the form $\Tr (X^n)$.
Expanding the exponential of the perturbation, we see
that the partition function of the perturbed model
can be computed by summing over correlators
in the Gaussian model, with insertions of powers of
$\Tr (X^n)$. At each order in  the coupling constant,
we have the correlator of a  multi-trace operator
in the Gaussian model, which as shown in section (\ref{OWP-3BPs}),
amounts to summing over Hurwitz classes with
three branch points. These  Hurwitz classes have a permutation
 $\sigma$ in the conjugacy class $ [n^m] $ coming from the
operator insertion, a permutation  $\g$ in the conjugacy class $[2^{nm\over 2}]$
 from the Wick contraction and  a permutation $ \tau $ which
is the product $  \gamma^{-1} \s^{-1} $.

\subsubsection{ Perturbation by  $ tr X^4 $ and Hurwitz numbers } 

For concreteness we start with perturbation by $ tr X^4$ 
\bea
Z ( X ) =  \int dX  e^{ { - N }  ( { 1\over 2 } tr X^2 + { g^2 \over 4}  tr X^4 )  }
\eea

The free two-point function is
\bea
&& \langle X^{i}_j X^k_l  \rangle_0  \cr
&& =   \int dX
 e^{ { - N \over 2 } tr X^2   }   X^{i}_j X^k_l  \cr
&& =  { 1 \over N } \delta^i_l \delta_j^k \cr
&& = { 1 \over N } { \bf \langle }   X^{i}_j X^k_l  { \bf \rangle }
\eea
This is the same as (\ref{bascor}) up to an immaterial factor of 
${ 1\over N } $.

Expanding the exponential
\bea
 Z ( X ) &=&   \sum_{ k } \int dX  e^{ -{N\over 2 } tr X^2 }
 { ( - N g^2 tr X^4 )^k \over 4^k k!  } \cr
 &=&  \sum_{ k } { (-g^{2})^{k} N^k  \over 4^k k! }
                         \langle   ( tr ( X^4   ) )^k  \rangle_0  \cr
 &=&   \sum_{ k } { (-g^{2})^{k} N^k  \over 4^k k! }
                         \langle tr_{ 4k} ( \sigma \bX   ) \rangle_0  
\eea 
We have used the observation from section \ref{revtst} 
that multi-traces can be written as a trace in a tensor 
product space with a permutation inserted. 
In this case, $ ( tr (X^4) )^k  $ can be written as a trace in 
$ V^{\otimes 4k } $, which is indicated by  the subscript 
in $ tr_{ 4k}$. The permutation $ \sigma $ has $k$ cycles of 
length $4$. So $ C_{\sigma } = k $. We can now write  
\bea\label{manipt4}  
 Z ( X ) 
& = & \sum_k  ~ { (-g^{2})^{k}  \over (4k)! } N^{ C_{ \s} }
~  { (4k)! \over 4^k k! }
                ~ \langle tr_{ 4k} ( \sigma \bX   ) \rangle_0  \cr
& = &  \sum_{ k}  { (-g^{2})^{k} \over ( 4k )! }  N^{ C_{ \s} }
\sum_{ \s \in  [ 4^k  ]  } ~ 
 \langle tr_{ 4k} ( \sigma \bX   ) \rangle_0  \cr
& =  & \sum_{ k} { (-g^{2})^{k} \over ( 4k )!}  N^{ C_{\s} - 2k  }
    \sum_{ \s \in [ 4^k ]  } \sum_{ \g \in  [2^k ]  }
 \sum_{  \tau \in S_{4k} }  tr_{4k}  ( \s \g ) \cr
& = &   \sum_{ k} { (-g^{2})^{k} \over (4k)!}
 \sum_{ \s \in [ 4^k ] } \sum_{ \g \in  [ 2^k ]  }
 \sum_{  \tau \in S_{4k} } \delta_{S_{4k}} ( \s \g \tau ) N^{ C_{ \s} + C_{ \tau } - 2k }\cr
& =  &  \sum_{ k } (-g^{2})^{k}  \sum_{ f  ( [4^k],[2^{2k}] )  :\Sigma_h \rightarrow \mP^1 } 
{ 1 \over { | Aut ( ~ f( [4^k],[2^{2k}]  )  } ~ ) |  }  N^{ 2 - 2h }
\label{quartic}
\eea
The factor $ | [ \s ] | $ we needed in (\ref{judicious}) arose naturally as a result
of expanding the standard normalization of the potential in the Matrix Model.

For any $g$ we have a Hurwitz interpretation of the Matrix model correlator. As $g$
approaches $g_c$ the partition functions diverges and the standard string interpretation
of the 90's emerges i.e CFT coupled to pure $ c \le 1 $ matter. In this case, it is just pure gravity.
 Very importantly now we also have an interpretation
in terms of {\bf continuum worldsheets and holomorphic maps
for any $g$}.

The double-scaled string theory (see the review \cite{dgz})
related to
the pure gravity arises in the limit,
\bea
 g \rightarrow g_c = -{ 1 \over 12 }
\eea
where it can be proved that
\bea
Z_h  \sim ( g_c - g )^{ 5 \chi_h \over 2 }
\eea
and where it becomes appropriate to
define a new genus counting parameter
\bea
\kappa^{-1}  = N ( g- g_c )^{ 5/4}
\eea
Our results imply that {\it  2D gravity coupled to different minimal models
arises from critical limits of generating functions of Hurwitz numbers. }

Fluctuations in the matrix model are of size ${1\over N^2}$. By switching to eigenvalue variables,
one can use a classical (saddle point) analysis to extract the leading large $N$ behavior. This will
give Hurwitz numbers for maps with both worldsheet and target a sphere. Let us make this explicit.
The large $N$ eigenvalue density for this model has been calculated \cite{BIPZ}
\bea
\rho ( \lambda ) = { 1 \over 2 \pi g } ( \lambda^2 + 1 + { 1 \over 2 } a^2 )
\sqrt  { ( a^2 - \lambda^2 ) }
\eea
where
$$ a^2 = {2\over 3}(-1+\sqrt{1+12 g})\, . $$
The free energy is given as usual by the log of the partition function $Z$. For the model we consider here
\bea
F = N^2 \Big( \int d \lambda d \mu \rho ( \lambda ) \rho ( \mu ) \log | \lambda - \mu |
- { 1 \over g } \int d \lambda \rho ( \lambda )
( { \lambda^2 \over 2 } + { \lambda^4 \over 4 } ) \Big)
\eea
Using the explicit eigenvalue density  the free energy
is evaluated  to obtain \cite{BIPZ}
$$
{F(g)-F(0)\over N^2} =
\sum_{k=1}^\infty g^{k}{(-1)^{k+1} (2k-1)!!6^k \over 2k\, (k+2)!}\, .
$$
This free energy can be recovered by summing the connected planar diagrams.
To make sure we pick the connected part in the delta function $\delta ( \s \g \tau )$
appearing in (\ref{quartic}), we keep only triples $\sigma$,$\g$,$\tau$ which are transitive.
To be more precise, the coefficient of $g^k$ in the free energy is the
number of times $\sigma\gamma = \tau^{-1}$ for (i) $\sigma$ summed over $[4^k]$ (ii) $\gamma$ summed over $[2^{2k}]$
and (iii) the group generated by $\sigma,\g,\t$ acts transitively on
 the set $\{ 1,2,...,4k\} $. In other words $\sigma,\g,\t$ generate the whole of $S_{4k}$.  
In what follows we use the notation $H^g_{\alpha,\beta}$ to denote the number of degree $d$
branched covers of $\mP^1$ by a genus $g$ connected Riemann surface with three
branch  points, the first two having ramification profiles $\alpha$ and $\beta$,
and the third having arbitrary ramification. If the cover has automorphism group
${\rm Aut} f_{\sigma,\gamma}$ it is counted with multiplicity $1/|{\rm Aut} f_{\sigma,\gamma}|$. This notation coincides with that of  \cite{GJV}.
The Hurwitz number $H^0_{[2^{2k}],[4^k]}$ is given by
the absolute value of the coefficient of $g^k$ in the free energy.
From the free energy above
\begin{align}
\fbox{
\vspace*{.6cm}
$~~~~~~~~~~~~~~~~~~ H^0_{[2^{2k}],[4^k]}~~ = ~~ {(2k-1)!! 6^k \over 2k (k+2)!} \,  ~~~~~~~~~~~~~~~~~~~~ $
\vspace*{.6cm}
}
\end{align}
In terms of delta functions 
\bea 
\sum_{ \sigma \in [4^k]  } \sum_{  \gamma \in [2^{2k}]  } ~   
\sum_{ \tau \in S_{ 4k} :   C_{\tau} = k+2    } { 1 \over (4k)! } 
\delta^{ ( conn :  0 )  } 
 ( \sigma \gamma \tau ) =   {(2k-1)!! 6^k \over 2k (k+2)!}
\eea 
The superscript $ conn $ on the delta functions indicating that we are
restricting to transitive triples which give rise to connected covers. 
The superscript $0$ indicates that we are restricting to genus zero 
worldsheet, which is equivalent (see (\ref{countgenus}))
to restricting  $  C_{ \tau } = 2k+2 $.

By expanding the partition function itself, we can obtain the Hurwitz numbers
$\tilde{H}^0_{[2^{2k}],[4^k]}$ which are defined as before
 except that the covers need not
be connected. Finally, correlators at large $N$ are given as moments of the
large $N$ eigenvalue density. For an arbitrary $2j$ cycle\footnote{We assume that $\sigma$ has a single cycle.
Using the large $N$ eigenvalue density one is only able to compute the leading contribution to any given correlator. This
is disconnected if $\sigma$ is not a single cycle.} $\sigma$ we have
$$
{1\over N^j} \int dX \, tr_{2j} (\sigma \bX  )e^{ { - N }  ( { 1\over 2 } tr X^2 + { g^2 \over 4N}  tr X^4 )  }
= \int_{-a}^a d\lambda\rho(\lambda )\lambda^{2j}
$$
Expanding this correlator in powers of $g^k$, the expansion coefficients are related to
$\tilde{H}^0_{[2^{2k+2j}],[\sigma\circ 4^k]}$.

\subsubsection{ Perturbation by $tr (X^3) $ and Hurwitz numbers }   
To obtain the Hurwitz numbers $H^0_{[2^{3k}],[3^{2k}]}$ we consider the matrix model
\bea
Z ( X ) =  \int dX e^{ { - N }  ( { 1\over 2 } tr X^2 + {g\over 3}  tr X^3 )  }\, .
\eea
For this model \cite{BIPZ},
$$
{F(g)-F(0)\over N^2}=-{1\over 2}\sum_{k=1}^\infty {(8 g^2)^k\over (k+2)!}{\Gamma (3k/2)\over\Gamma (k/2+1)} \, ,
$$
so that
\begin{align}
\fbox{
$~~~~~~~~~~~~~~
H^0_{[2^{3k}],[3^{2k}]}~~=~~{1\over 2} {8^k\over (k+2)!}{\Gamma (3k/2)\over
\Gamma (k/2+1)}\, .
~~~~~~~~~~~~~~$
}
\end{align}

In terms of delta functions 
\bea 
\sum_{ \sigma \in [3^{2k} ]  } \sum_{  \gamma \in [2^{3k}]  } ~   
\sum_{ \tau \in S_{ 6k } :   C_{\tau} = k+2    } { 1 \over (6k)! } 
\delta^{ ( conn :  0 )  } 
 ( \sigma \gamma \tau ) =   {1\over 2} {8^k\over (k+2)!}{\Gamma (3k/2)\over
\Gamma (k/2+1)}
\eea
The connectedness condition is a transitivity constraint
on the triples, the genus zero condition is equivalent 
to $ C_{\tau} = 2 + k $.

\subsubsection{ Perturbation by $tr X^6$ and Hurwitz numbers   } 
Finally, we consider the matrix model with $tr X^6$ potential
\bea
Z ( X ) =  \int dX  e^{ { - N }  ( { 1\over 2 } tr X^2 +  {g \over 6 } tr X^6 )  }\, .
\eea
The large $N$ limit for this model has been studied in \cite{Cicuta}. From \cite{Cicuta}
we obtain the free energy
$$
F = { - a^4 \over 12}  + { 7 \over 12  } a^2 - { 1\over 2 } \log a^2
$$
where
$$
\hat g a^6 + a^2 -1 = 0\qquad \hat g = 60 g
$$
Using these two equations we will now develop the series expansion of the free energy.
This last cubic is easily solved to give
\bea\label{eqa2}
a^2 =
 { 1 \over \sqrt { 3\hg} }
 \bigl( \sqrt { 1 + { 27 \hg \over 4 } }  + \sqrt { 27 \hg \over 4 } \bigr )^{ 1 \over 3 }
 -  { 1 \over \sqrt { 3\hg} }
 \bigl ( \sqrt {  1 + { 27 \hg \over 4 } }  - \sqrt { 27 \hg \over 4 }  \bigr)^{1 \over 3 }
\eea
We know that we have the correct root because it is clear that $a^2$ has an expansion starting at $1$ for small $\hg$.
It is now easy to obtain the following series expansions
\bea
a^2 = \sum_{ k=0 }^{ \infty } { 3k \choose k } {1 \over {2k+1 } } (-1)^k \hg^k
\eea
$$
a^4 =\sum_{k=0}^\infty {(3k+1)!\over (2k+1)!(k+1)!}(-1)^k \hg^k
$$
and
$$
\log (a^2)=\sum_{k=1}^\infty (-1)^k {(3k-1)!\over k! (2k)!}\hg^k
$$
Collecting these results we obtain
\be
{F(g)-F(0)\over N^2}=
\sum_{k=1}^\infty {(-1)^{k+1}\over 2}{(10)^k (3k-1)!\over (2k+1)!(k+1)!} g^k
\label{freeenergy}
\ee
and hence the Hurwitz numbers
\begin{align}
\fbox{$
~~~
~~  ~~ H^0_{[2^{3k}],[6^{k}]} ~~=~~ {1\over 2}{(10)^k (3k-1)!\over (2k+1)!(k+1)!}\, ~
$
}
\end{align}

To check our result (\ref{freeenergy}), we will now explain how the free energy, or equivalently, Hurwitz numbers,
can be computed from the class algebra coefficients of the symmetric group.
Return to the formula (\ref{basiccorr2})
\bea\label{corrsym}
&& \langle  tr_{2n} ( \sigma X ) \rangle = { 1 \over  |[\s]| }
\sum_{\rho\in [\s]}\sum_{\gamma\in [2^n]}\sum_{\tau\in S_{2n}}
N^{C_\tau}\delta_{S_{2n}}(\rho\gamma\tau)\cr
&&  =   \sum_{[\tau]}f_{ \s , \g}^{ \tau } { Sym ( [\s] ) \over Sym ( [ \tau ] )  }
\eea
where $f_{ \s , \g}^{ \tau }$ are the class algebra coefficients
$$ T_\s T_\g = f_{ \s , \g}^{ \tau }T_\tau $$
In this last expression, $T_\psi$ stands for the sum of elements in the conjugacy class $[\psi]$.
The class algebra coefficients are easily evaluated with the help, for example, of the Symmetrica program \cite{symmetrica}.

\begin{figure}
\begin{center}
\resizebox{!}{2.5cm}{\includegraphics{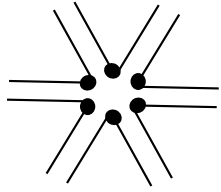}}
\caption{The graphical representation of $\tr (X^6)$.}
\label{fig:opp}
\end{center}
\end{figure}

The partition function is
$$
Z=1-{gN\over 6}\langle\tr (X^6)\rangle_0 +{1\over 2!} \left({gN\over 6}\right)^2\langle\tr (X^6)^2\rangle_0 + ...
$$
To evaluate $\langle\tr (X^6)\rangle_0$ we need to consider the product
\be
[\sigma][\gamma]=[6][2^3]=6[3\, 1^3]+8[2^2 1^2]+5[5\, 1]+4[4\, 2]+3[3^2]
\label{grpalg}
\ee
The first two terms on the right hand side each correspond to cycles with 4 parts ($C_\tau =4$) so these give the
leading (planar) contribution. The remaining four terms give a torus (down by ${1\over N^2}$) correction.
It is in fact straightforward to identify each of the terms above with a particular double line diagram. Our original
operator $\tr (X^6)$ is a sum over 6 indices. We will represent each index by a dot, as shown in  Figure \ref{fig:opp}.
These are the object that are permuted by $S_6$.

To obtain the double line diagram, the 6 ``stubs'' must be connected by fat (double line) propagators. One such connection is
shown in \ref{fig:frstrbbn} below. The connected double line diagram is in fact a graphical representation of $\tau$.
To read $\tau$ from this double line diagram, recall that each dot is an object that gets permuted.
Each closed loop in the double line diagram will contain at least one dot. These loops tells you how the dots are permuted by the
action of $S_6$. Thus a loop with $n$-dots in it corresponds to a $n$-cycle in $\tau$. For the double line diagram in Figure \ref{fig:frstrbbn}
it is clear that $\tau =[1^3 3]$. From (\ref{grpalg}) we see that the relevant group algebra coefficient is 6 and hence
this diagram has a coefficient

$$
f^{[1^3 3]}_{[6][2^3]}{ Sym ( [\s] ) \over Sym ( [ \tau ] )  }=6 {6\over 3\cdot 3!}=2
$$

\begin{figure}
\begin{center}
\resizebox{!}{2.5cm}{\includegraphics{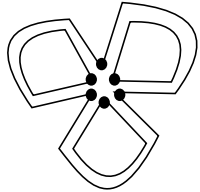}}
\caption{For this double-line diagram $\tau=[1^3 3]$.}
\label{fig:frstrbbn}
\end{center}
\end{figure}

\begin{figure}
\begin{center}
\resizebox{!}{2.5cm}{\includegraphics{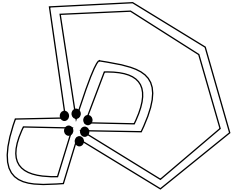}}
\caption{For this double-line diagram $\tau=[1^2 2^2]$.}
\label{fig:scndrbbn}
\end{center}
\end{figure}

For the double line diagram in Figure \ref{fig:scndrbbn}
it is clear that $\tau =[1^2 2^2]$ and hence
this diagram has a coefficient

$$
f^{[1^2 2^2]}_{[6][2^3]}{ Sym ( [\s] ) \over Sym ( [ \tau ] )  }=8 {6\over 2^2 2! \cdot 1^2 2!}=3
$$

\begin{figure}
\begin{center}
\resizebox{!}{2.5cm}{\includegraphics{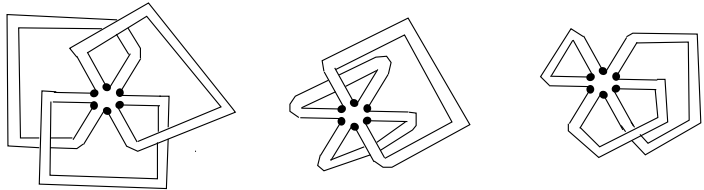}}
\caption{Non-planar double-line diagrams contributing to $\langle \tr (X^6)\rangle $}
\label{fig:lstrbbn}
\end{center}
\end{figure}

The non-planar double line diagrams are shown in Figure \ref{fig:lstrbbn}. They correspond to
$\tau=[3\, 3]$, $\tau=[4\, 2]$ and $\tau=[1\, 5]$ respectively. Notice that each term in
(\ref{grpalg}) corresponds to a unique double line diagram. At higher orders in $g$ this is no
longer the case - there may be two different double line diagrams with the same cycle structure
for $\tau$.

It is now simple to obtain
$$
Z=1-{g\over 6}(5N^4+10N^2)+{g^2\over 36\cdot 2!}(700 N^6 + O(N^4))+ O(g^3)
$$
The free energy is obtained by summing connected diagrams.
To get the connected diagrams, take the log
$$
\log (Z) = -{g\over 6}(5N^4+10N^2)+{g^2\over 36\cdot 2!}(700 N^6 + ...)+... -{1\over 2}
(-{g\over 6}(5N^4+10N^2)+{g^2\over 36\cdot 2!}(700 N^6 + ...)+...)^2 +...
$$
$$ = -5{gN^4\over 6} + 25 {g^2 N^6\over 3} + ...$$
This matches the free energy computation above.

We can state the result in terms of the delta functions 
on symmetric groups 
\bea 
   \sum_{ \rho \in [6^{k} ] \in S_{6k } }  
\sum_{ \gamma  \in [2^{3 k } ] \in S_{6k  } } 
\sum_{ \tau \in S_{6k} : C_{ \tau } = 2 + 2k  } 
 { 1 \over (6k)! } \delta^{ { (conn ; h=0 ) } }   ( \rho ~  \gamma ~  \tau )  
= {1\over 2}{(10)^k (3k-1)!\over (2k+1)!(k+1)!}
\eea 
The superscripts on the delta function indicate that 
we are restricting to genus zero connected maps. 
The constraint of the worldsheet being connected 
is implemented by requiring that the $ \sigma ,  \gamma , \tau  $ 
generate the whole symmetric group $S_{6k} $.

\subsection{Universal expressions, Branes and Resolvents}\label{universal}

Having shown (section (\ref{OWP-3BPs})) that the 1-point function of any 
multi-trace operator in the Gaussian model counts holomorphic maps with three 
branch points, it is clear that the partition function and correlation
 functions of the Gaussian model perturbed by a general potential 
can likewise be expressed in terms of holomorphic map counting 
with three branch points. It suffices to 
just treat the exponential of the perturbation as 
an observable in the Gaussian model. Indeed we have seen 
explicit examples  in section (\ref{newhurwitz}). 
It is possible to express this  simple consideration in  elegant formulae 
if we introduce some definitions. 

Given two 
 permutation groups  $ S_{n_1} $ and $S_{n_2}$, 
we can define a canonical embedding of $ S_{n_1} \times S_{n_2} $ 
into $S_{n_1 +n_2}$, by considering $S_{n_1} $ as 
permutations of $ \{ 1 \cdots n_1 \}  $ and $S_{n_2} $ as permutations 
of $ \{ n_1 + 1 , \cdots , n_1 + n_2 \} $. Given two permutations 
$ \s_1 , \s_2 $ respectively in $ S_{n_1} $ and $S_{n_2}$ we 
write $ \s_1 \circ \s_2 $. We will call this the outer product
of the two permutations.  This operation can clearly be extended 
to an arbitrary string of permutation groups. 

We can define an {\it outer-exponential} of a permutation $\sigma \in S_{i} $  
as follows 
\bea 
{\bfe}_{\circ}^{ \s } = \sum_{k=0}^{\infty } { 1 \over k!}  \s^{\circ k } 
\eea 
where $~~ \s^{\circ k } \equiv  \s \circ \s \circ \cdots \circ \s ~~ $
is using the embedding of $S_{i}^{\times k } $  in 
$S_{ki} $. The definition can be extended, by linearity and 
 distributivity of the outer product over sums,     to the outer 
exponential of a 
sum of permutations.   
\bea\label{expandexp}  
\bfe_{\circ}^{ \sum_i g_i \s_i  }
= \sum_{k=0}^{\infty } 
  \sum_{ i_1 , i_2 \cdots , i_k } { g_{i_1} g_{i_2} \cdots g_{i_k } \over k! } 
                    ~~  \s_{i_1} \circ \s_{i_2} \circ  \cdots \circ \s_{i_k} 
\eea 

Another useful definition will be a generalization of 
the delta functions for multiple permutations in the same symmetric group 
(\ref{delsd})
to a delta function,  denoted $\bdlt $, which takes three arguments in  
symmetric groups $S_{d_1}, S_{d_2} , S_{d_3} $ of permutations 
of $ \{ 1 , 2 , \cdots  , d_1 \}, \{1 , 2 , \cdots  , d_2 \}$ and   
$\{1 , 2 , \cdots  , d_3 \} $ where the $d_i$ are arbitrary 
positive integers.  If 
the degrees are not equal the delta function is defined to be zero. 
When the degrees are equal the $\bdlt $ is defined as equal 
to the usual $ \delta_{ S_{d} } $.  In other words it is equal to 
$1$ if the three permutations are in the same $S_{d} $
and multiply to $1$.

Calculations similar to those involved in (\ref{judicious}) and  
(\ref{manipt4})
lead in the case of a perturbation of the Gaussian term  by $g\Tr X^i$ 
as 
\bea 
Z  && = \boldsymbol{ \delta } ~ ( ~   \bfe_{\circ} ^{g~c_i  } ~ { \bf T } ~   { \bf \Omega } ~ ) 
\eea  
Here $c_i $ is a cyclic permutation of length $i$ in $S_i $. 
$  { \bf T }$ is a sum of permutations in $[2^{p}]  $ 
which is itself summed over $p$. As $p$ increases 
 $S_{2p}$ viewed as permutations of $\{ 1 , 2 , \cdots , 2p \}$, 
which can be viewed as a subgroup of $S_{\infty} $.  
\bea 
 { \bf T } = \sum_{p } \sum_{ \gamma  \in [2^p] \in S_{2p}  } \gamma  
\eea 
Similarly 
\bea 
  { \bf \Omega }  = \sum_{ p } \sum_{ \tau \in S_p } 
 N^{C_{ \tau }}  \tau 
\eea 
Keeping track of all the factors of $N$ from normalizations of 
the perturbations and 2-point function 
as done in  the previous examples shows that 
the power of $N$ is consistent with a string interpretation 
according to the Riemann-Hurwitz formula. Likewise, the 
combinatoric factors lead to $ { 1 \over |Aut f | } $.

A general potential $ tr(X^2)+V(X ) $ is naturally associated with a 
sequence of formal sums of permutations. 
Usually one considers single trace terms in the potential, 
so that the $ \hat V = \sum_i g_i c_i $ is a sum over single cycles.
For multi-trace perturbations we have a sum of more general 
permutations and $  \hat V = \sum_i g_i \s_i $. 
In this case the partition function is 
\bea  
Z = \langle 1 \rangle_{ tr X^2 + V } 
  = \boldsymbol{ \delta } (~  \bfe_{\circ}^ { \hat V }~   { \bf T }  ~ { \bf \Omega } ~ ) 
\eea 
The expansion of the exponential according to (\ref{expandexp})
contains terms in permutation groups of different degrees. 
For all the terms of a fixed degree $d$,  
the definition of $ \bdlt $
 picks out from the sums in $ { \bf T } $ and $ { \bf \Omega } $  
precisely those terms which belong to symmetric groups of 
the same degree $d$. Then $ \bdlt $ reduces to 
a sum over  $d$ of terms $ \delta_{S_{d} } $.

This extended language of $ \bdlt , { \bf T } , { \bf \Omega } $
allows us to give neat expressions for some 
key quantities in Matrix theory. Consider the function 
$ {\rm Det} ( x - X )$. 
Using the earlier manipulations we can write 
\bea\label{bakakh}  
\langle ~  {\rm Det} ( x - X )  ~ \rangle_{ tr X^2 + V }
 ~ = ~  \boldsymbol{ \delta } ( ~   (D\circ \bfe_{\circ}^ { \hat V }) ~  { \bf T }~ { \bf \Omega } ~  ) 
\eea
where $ D = \sum_i x^{N-i} (-1)^i  \sum_{ \sigma \in S_i } (-1)^{\sigma}$. 
This uses the expansion of the determinant
\bea 
\langle ~ { \rm Det } ( x - X ) ~  \rangle
= \sum_i x^{N-i} (-1)^i  \langle ~  \chi_{ [1^{i} ]} ( X )  ~ \rangle  
\eea 
where $ \chi_{ [1^{i} ]} ( X ) $ is the Schur polynomial 
for the representation of $U(N)$ corresponding to the 
anti-symmetric Young diagram which is a column of length $i$
or equivalently $i$ rows of length $1$ as indicated by $[1^i]$. 
The Schur polynomial has an expansion 
\bea 
\chi_{[1^i]  } ( X ) = 
\sum_{ \sigma \in S_i } { (-1)^{ \sigma }\over i ! }  \Tr_i ( \sigma \bX ) 
\eea 
where $ (-1)^{\sigma}$ is $1$ if the permutation is even 
and $-1$ if it is odd.

The 1-point function  $\langle  {\rm Det} ( x - X ) \rangle  $ 
defines the   Baker-Akhiezer function which 
is used in \cite{mmss} to argue that the target space of the 
1-Matrix model is 
$\mP^1$.
 This conclusion is reached by identifying
 the insertion of the exponentiated macroscopic loop
operator ${\rm exp}(Tr\log (x-X))= {\rm Det } ( x - X )$ as the matrix model 
description of an FZZT brane \cite{bdss}.
Since $x$ describes the position of the brane, that is,
 the allowed places where open strings
can end, it is  natural to identify $x$ with the target
 space coordinate.
Since the Baker-Akhiezer function is an entire function, $x$ runs over $\mP^1$.
We can write formally 
\bea\label{BK}  
 {\rm Det} ( x - X ) \rightarrow  D = 
 \sum_i x^{N-i} (-1)^i  \sum_{ \sigma \in S_i } (-1)^{\sigma} \sigma 
\eea  
This  a universal formula 
which works for correlation functions of  $ {\rm Det} ( x- X )$ 
inside the delta functions as in (\ref{bakakh}).  
 The lesson which emerges from  equations (\ref{BK})  and (\ref{bakakh}) is 
that the operator which creates  FZZT branes 
is {\it a fermionic condensate of worldsheet ramification points }. 
It would be interesting to clarify the connection 
between this remark and the interpretation  of FZZT-branes
as a condensate  of long strings \cite{condlong}.

For the resolvent
\bea 
R  ( x ) = \Tr {  1 \over x - X  }    = 
 \sum_{ k=0}^{\infty } x^{-k-1} \Tr X^k  
\eea 
we have
\bea 
\langle  R (x)   \rangle_{ tr X^2 + V }
= \boldsymbol{ \delta } ( ~  ( R \circ e_{\circ}^ { \hat V })  ~  { \bf T } ~ { \bf \Omega }~  ) 
\eea 
where $ R = \sum_{ k} x^{-k-1} c_k  $.
The resolvent is a very useful auxiliary function
 used when determining the large $N$ eigenvalue 
density. One can define a quadratic equation which determines the resolvent.
This equation defines
the spectral curve, which has a natural interpretation in terms of topological string theory
on certain Calabi-Yau manifolds \cite{Marino:2004eq,Dijkgraaf:2002fc}.
The spectral curve can also
be used to compute correlators at genus zero and higher genus \cite{Ambjorn:1990ji}. 

It is interesting that the area dependence in the 
 string theory  of 2dYM \cite{gt}
appears through 
  exponentials such as $ e^{A T^{(d) }_2}$, where $T_2^{(d)} $ is the 
sum of elements in the conjugacy class of simple transpositions in $S_d$. 
At each order in the expansion of the exponential, 
the product $ ( T_2^{(d)})^k $  is a  product
in the class algebra of a fixed symmetric group $S_d$. 
These products lead to counting problems with 
$k$ simple branch points. For the 1-matrix model at hand we 
encounter the outer exponential of (\ref{expandexp}). In 
a sense the difference between the areas in  string theory of 
2dYM (generalized ``areas''  couple to higher branch points  
\cite{cmrii,highercas}), which sums over different numbers of 
branch points,  and  the couplings of the 
string theory of the 1-matrix model, which is 
about three branch points but different ramification types,
 arises from the choice of exponentials
in symmetric groups.

\section{ Feynman Graphs  and the absolute Galois group }\label{FeynmanGalois}

Traditionally we think of the Feynman graphs for matrix model correlators
in terms of { \it double line diagrams}, following 't Hooft \cite{thooftplanar}.
Equivalently we can use the language of {\it ribbon graphs},
where the propagators are single lines, but each vertex is
equipped with a cyclic ordering of the edges (see for example
\cite{Looijenga}). In section \ref{onemathurwitz} we have expressed
correlators using triples  of permutations (which multiply to one) of the kind that appear
in counting branched covers with 3 branch points. This allows an immediate
use of a theorem of Belyi to deduce an interesting connection to curves
and maps defined over algebraic numbers (section (\ref{3tobarQ})).
Grothendieck exploited this theorem of Belyi to show that
the absolute Galois group $ Gal ( \bmQ / \mQ ) $ acts on
the equivalence classes of triples of permutations (Hurwitz-classes for
branched covers with 3 branch points). In his discussion, Grothendieck
used the notion of Dessin D'Enfants (Dessins for short), 
which are yet another way
to talk about ribbon graphs or double-line Feynman diagrams.
The upshot is that ``double-line Feynman diagrams of the 1-Matrix Model'', 
``Dessins '' , ``ribbon graphs'' , ``triples of permutations'' 
are all different descriptions of 
 the same thing, and they admit an action of the absolute Galois group.

Having a precise characterization of Feynman diagrams 
in terms of triples of permutations allows us then to
observe that the Galois group acts on Feynman graphs of the one-matrix model,
or equivalently on the pairs $ (\Sigma_h , f )$ where $ \Sigma_h $ 
is the string worldsheet and $f$ a holomorphic  map
 to the target space $ \mP^1$ with 3 branch points. 
 Most of the mathematical elements of this remark
are found already in \cite{Looijenga,baueritzyk}, but have not
been fully interpreted and exploited in the string theory literature.
The current  section explains, with concrete examples, 
 the role of the Galois group 
in organising the Feynman graphs of the 1-Matrix model. 
In section \ref{MMCED}, we will extend our considerations to  
 multi-matrix models, where we
are lead to introduce a new combinatoric object, colored-edge Dessins.
In section \ref{CEDGI} we will use  these to find new results on invariants
of the Galois action on ordinary Dessins.

\subsection{ 3 Branch points :  The Observable, The Wick contraction and the
Product }

The key result from the previous section (\ref{OWP-3BPs})
 is that {\it any} correlator of
the one matrix model, Gaussian or perturbed,
 can be written as a sum over maps from a Riemann
surface to the sphere. The genus $h$ of the Riemann surface determines the
power of $N$, which is $N^{2-2h}$,  at which it contributes.
 The map has three branch points with monodromies $\s$,
$\g\in [2^n]$ and $\t$. The monodromy $\s$ determines the observable whose correlator
we are computing, $\g$ is determined by the Wick contraction 
and $\t$ is the product $\g^{-1}\s^{-1}$. Our goal in this section is to explore and
develop the consequences of this observation.
We will associate
$\sigma$ with the ramification over 0,
$\gamma$ with the ramification over 1 and
$\tau$ with the ramification over $\infty$. To reflect this in our notation
we will sometimes refer to $\s,\g,\t$ as $\s_0,\s_1,\s_\infty$.

\subsection{ Belyi's theorem  : From Three branch points to $ \bmQ $  }\label{3tobarQ}

Useful references for this section are \cite{schneps,lanzvon,Joubert}.
 In what follows we will consider
algebraic number fields, which are field extensions of the field of rational numbers $\mQ$.
Thus, an algebraic number field is a field that contains $\mQ$ and has finite
dimension when considered as a vector space over $\mQ$.
The field $\bmQ$ obtained by adding all algebraic numbers to $\mQ$ will play an important
role in what follows.

A classic theorem due to Weil states that a curve is defined over $\mathbb {\overline  Q }$
if there exists a non-constant holomorphic function 
$f : \Sigma_h  \rightarrow {\mathbb P}^1 {\mbC}$
all of whose critical values lie in $\mathbb {\overline  Q }$.
Belyi's theorem on algebraic curves states that
given any algebraic curve defined over $ \mbC$, 
it can be defined over $ \bmQ$ if and only if 
there exists a holomorphic function $ f : \Sigma_h \rightarrow {\mathbb P}^1  $
such that its branch points lie in the set $\{  0, 1 , \infty \} $. 
Following conventions in Belyi literature, we will use the 
notation $X = \Sigma_h $ and $ \beta = f $. 
The pair $(X,\b)$ where $X$ is a compact Riemann surface
and $\b : X \to {\mathbb P}^1$ is a holomorphic map unbranched 
outside the set  $\{ 0,1,\infty\}$, both defined over $ \bmQ$, 
 is called a Belyi
pair and we call $\b$ a Belyi function.

The inverse image under $\beta$ of the closed
interval $[0,1]$ defines a Grothendieck
Dessin (or just Dessin for short). The points in the preimage 
of $0$ are marked with a
black vertex and the points in the preimage of $1$ are marked with
a white vertex. Permutations $ \sigma_0 , \sigma_1$ 
can be assigned to Dessins by labelling the edges 
and going round the black and white vertices respectively
(see for example the explanation in \cite{Joubert}).

A clean Belyi map is one which has all ramification orders equal to
 2 over the point at $1$.
The Dessin corresponding to a clean Belyi map has exactly two edges
 for every preimage of $1$.
Given a general map $\alpha$ of Belyi type, we can get
a new Belyi map $\beta$ with $ \beta = 4 \alpha ( 1  - \alpha ) $,
such that $ \beta $ is a clean Belyi map. 
Recall from the previous subsection that since $\g\in [2^n]$,
it is clear that it is the clean Belyi maps that arise as Feyman graphs in the
1-matrix model. The ramification orders above the 
point $0$ are described by $\s$.

\begin{figure}
\begin{center}
\resizebox{!}{4.0cm}{\includegraphics{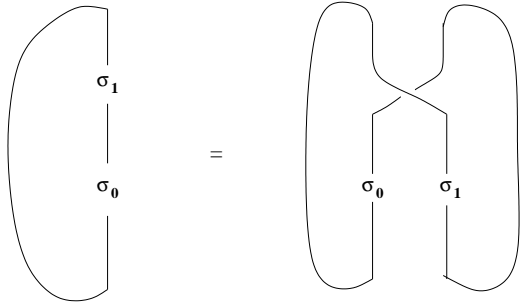}}
\caption{ Tensor operator equality corresponding to cleaning of Belyi map   }
\label{fig:tenscleaning}
\end{center}
\end{figure}

From the point of view of a Dessin,
the process of cleaning amounts to converting 
the white vertices into black vertices 
and introducing white vertices in the middle of the 
edges joining the white vertices. If, with a 
labelling of the edges,  $ \sigma_0 \in S_d $ describes the 
permutation of edges around the black vertices, 
and $ \sigma_1\in S_d $ describes the permutation of the edges 
around the white vertices, then after the cleaning 
operation, $ (  \sigma_0 \circ \sigma_1 ) \in S_{2d} $ 
describes the permutation of the black vertices 
and a permutation which maps $ \{ 1 \cdots  d \} $ to $ \{ d+1 , d+2 , \cdots ,  2d \}  $ 
pairwise, i.e the permutation with cycle decomposition 
$ ( 1 ~ d+1 ) (2 ~ d+2) \cdots ( d ~ 2d ) $, describes 
the permutation around the white vertices. In terms of tensor 
diagrammatics, sequences $ \sigma_0 , \sigma_1$ acting on $ V^{\otimes d } $ 
have a trace given by the diagram on the left of Figure 
\ref{fig:tenscleaning}. 
The same trace can be described as a trace of something in 
$ V^{\otimes 2d } $ by a simple diagrammatic manipulation, 
as in the right of Figure \ref{fig:tenscleaning}.

As a side-remark, consider the Dessin associated to a clean Belyi pair. 
After choosing a marking of the Dessins, 
the oriented cartographic group $C_2^+$, an infinite discrete group, 
permutes these marked Dessins.
It is possible to prove that Dessins associated to a clean Belyi pair
(which are also the Feynman graphs in the 1-matrix model) are in
 1-1 correspondence with the
conjugacy classes of subgroups of  $C_2^{+}$ \cite{schneps}. 
The generators of $ C_2^+ $ are $ \rho_0 , \rho_1 , \rho_2 $
with relations  $ \rho_1^2 = 1 , \rho_0 \rho_1 \rho_2 =1 $.
Thus, Feynman graphs are in 1-1 correspondence
with conjugacy classes of subgroups of the Cartographic group.

\subsection{ $ Gal ( \bmQ / \mQ ) $  for organizing Feynman graphs    }\label{fromFeynDess}

An important group in number theory, called the
absolute Galois group $ Gal ( \bmQ / \mQ ) $, organizes all the
key properties of the algebraic numbers.
It is the group of all automorphisms of the algebraic closure  $\bmQ$ that fix
$\mQ$. By allowing the absolute Galois group to act on the numerical coefficients
appearing in the Belyi pair, we get an action of the group on Dessins.
The Galois group acts faithfully on Dessins\footnote{In fact, $ Gal ( \bmQ / \mQ ) $
acts faithfully on the set of genus 1 Dessins, on the set of genus 0 Dessins and even
on the set of trees\cite{schneps}.}. This means that the Galois group
acts faithfully on the Feynman diagrams, and sets of Feynman diagrams
can be assembled into Galois orbits.
We have exploited the symmetric groups for organizing the
Feynman graphs contributing to the Matrix model correlator.
Now we are saying that the absolute Galois group can
further be used to organize the Feynman graphs
into orbits.

The correlator $ \langle \cO_{ \s_0 } \rangle $
in the the Gaussian  1-Matrix model is a sum
over Hurwitz classes weighted by  $ 1/ |Aut f|  $.
Choosing a multitrace operator is a choice of $ [ \sigma_0 ] $.
Choosing a Wick contraction is a choice of
a permutation  $ \sigma_1$ from the conjugacy class $ [ \sigma_1 ] = [2^n ] $.
Then we sum over  $ [\sigma_\infty ] $ which runs over conjugacy 
classes that can appear in the product of permutations from  $[ \sigma_0 ]$ 
and $ [\sigma_1]$. When computing the correlator
of the gauge invariant operator, contributions are weighted by
$ { 1\over |Aut f|  } $ (see \ref{corr3point}).  
 It is known that the data  $ [\s_0] , [\s_1] , [\s_\infty] , Aut f  $ 
 are Galois invariants \cite{jost}.
  This means that every 1-point function $\langle \cO_{ \s_0 } \rangle$
is a sum over Galois invariant data.
The set of Hurwitz classes which share the above data
can be one or many, and they can fit in one or more 
complete Galois orbits. When there are multiple orbits for fixed
$ [ \s_0 ] , [ \s_1 ] , [ \s_\infty ] , Aut f $,
there will be a list of finite subgroups of $ Gal ( \bmQ / \mQ ) $
which will each act transitively on each orbit. 
In one of our examples of \ref{sec:examples}, this  Galois group will 
be an $S_3$ acting as  permutations of the three roots of (\ref{fod}).  

It is worth noting that  the Galois invariance of the conjugacy classes
$ [\s_0] , [\s_1] , [\s_\infty] $ is also  useful in proving the
finiteness of the length of the Galois orbit for any Dessin.
There is a lot of mathematical interest in 
developing a complete list of Galois invariants 
which can be used to determine when a pair 
of Dessins are in the same orbit and when they are not
\cite{schneps,lanzvon,GGA2,edschnepsI}.

In the above, we have described the route from Feynman graphs to
triples of permutations. Grothendieck relates these triples 
to Dessins. Now we explain how to obtain  the Dessins
directly from the Wick contractions without
going through the triple of permutations:
Each trace operator $\tr X^k$ corresponds to  a black vertex
with $k$ edges emerging from it, cyclically ordered
using the orientation on a plane.
Join the edges emerging from the vertices, in pairs, according to the
Wick contractions.
Insert a white vertex along every propagator and introduce extra handles
to avoid intersections.
The result is a clean Dessin.

We know how to get the Dessin from the Belyi pair:
the Dessin is the inverse image of the closed interval $[0,1]$ with points
in the preimage of 1 marked with a white vertex and points in the preimage of
0 marked with a black vertex. It is possible to go in the other direction and
obtain the Belyi pair from the Dessin \cite{desstriang,Joubert}. Start by placing a point 
within each closed region of the Dessin and label it as $\infty$. Connect this new point to the black 
and white points forming the boundary of the region, connecting multiple times to the same black or 
white point if it appears multiple times on the boundary of the region. The result is a set of triangles 
each of which has three vertices, one labeled 0 (for the black point), one labeled 1 (for the white point) and 
$\infty$ (the new point). Each triangle is a half-plane. If the triangle has  0, 1, and $\infty$ in counterclockwise 
order it is an upper half-plane and if not, it is a lower half-plane. Adjacent pairs of triangles can now be 
glued together along the shared portion of their boundaries. The result is a Riemann surface. It can be mapped 
to the Riemann sphere by using the identity map within each half-plane 
so that we have indeed obtained a Belyi pair.

The role of $ Gal ( \bmQ/ \mQ ) $ in organizing 
Feynman graphs of the 1-matrix model is rather different from the action of symmetries
we are used to in quantum field theory.
 Usually we relate correlators of different observables,
when the observables fall in representations of a symmetry group.
Another good example are the Schwinger-Dyson equations in the 1-matrix model,
which are a consequence of the invariance of the matrix integral under
changes of variable. The Schwinger-Dyson equations relate the correlators of different
observables to each other. In this case, the Galois group
$ Gal ( \bmQ / \mQ ) $ relates different contributions to a fixed observable.
Perhaps the MHV re-organization of Feynman diagrams \cite{wittenMHV} 
is a reasonable analogy to this organization, although we are not
aware of a group which relates the set of different Feynman
diagrams which are collected together in the MHV method.

\subsection{Examples}\label{sec:examples} 

The connections developed above can be made very concrete in the context of specific examples.
For a sequence of observables in the Gaussian 1-matrix model,
described by conjugacy classes $ [\sigma_0]$ in $ S_{2n}$, we will 
consider the class algebra multiplication $ [\sigma_0]\cdot [2^n] $. 
For each $ [\sigma_{\infty} ] $ appearing in the product, we will consider 

\begin{itemize}

\item
The Hurwitz classes, their genus and automorphism group

\item
The size of the Galois orbit

\item
The Belyi pair

\item
The field of definition

\end{itemize}

These choices deserve a few comments.
The nature of the Galois orbit will provide
insight into the symmetry which organizes the Feynman
graphs. The field of definition is interesting, since its giving
detailed information about the Belyi map itself.
Related to fields of definition is the moduli field  $K$ which is 
the intersection of all the fields of definition \cite{lanzvon}. 
The moduli field $K$ is interesting because $ {\rm deg} [ K : \mQ ] $ is equal to
the size of the orbit. In our examples, the need to 
distinguish $K$ from the minimal  field of definition 
will not be necessary.

\begin{itemize}

\item $ \langle tr X^2 \rangle $ : $[2] \cdot [2] = 2 [1^2]$ \\
 There is a single Hurwitz class,
 a single ribbon graph corresponding to $[1^2]$.
 Since there is a one-to-one correspondence between ribbon graphs
 and Dessins, this immediately
 implies that the size of the Galois orbit has to be $1$. Hence ${\rm deg} [ K : \mQ ] = 1 $,
 which means the field of definition is $ K = \mQ $. $|Aut (f)| = 2$.
 The  Belyi map corresponding to this observable is
\bea\label{examp1} 
  w={z^2\over z-{1\over 4}}
\eea 
 The corresponding Dessin is given in Figure \ref{fig:gdessin2}.

\item $ \langle (tr X)^2 \rangle $ : $[1^2] \cdot [2] =  [2]$ \\
 There is a single Hurwitz class, a single ribbon graph
 which immediately
 implies that the size of the Galois orbit has to be $1$. Hence ${\rm deg} [ K : \mQ ] = 1 $,
 which again means the field of definition is $ K = \mQ $. $|Aut ( f )| = 2$.
 From Figure \ref{fig:gdessin2} it is clear that the Dessin for this observable can be obtained by
 cleaning the Dessin whose Belyi map is $w=z$, which gives
\bea\label{examp2} 
 w=4z(1-z) 
\eea 

\begin{figure}
\begin{center}
\resizebox{!}{3.5cm}{\includegraphics{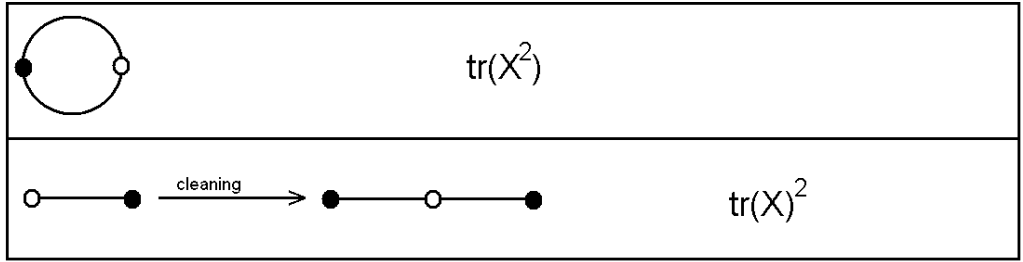}}
\caption{The Dessins corresponding to operators built using two matrices}
\label{fig:gdessin2}
\end{center}
\end{figure}

\item $ \langle tr X^4 \rangle $ : $[4]\cdot [2^2] = 2[2\, 1^2]+[4]$\\
 There are two Hurwitz classes, one for $[2\, 1^2]$ and one for $[4]$. One corresponds to a map from
 the sphere to the sphere and one to a map from the torus to the sphere. Since genus
 is a Galois invariant, this correlator
 gets contributions from two Galois orbits each of which have size $1$. Thus, the field of
 definition for both is again $ K = \mQ $. The map for the Dessin of genus zero is
\bea 
 w={(z-{3\over 2})^4\over (z-1)(z-2)} 
\eea 
 The two poles in the Belyi function are needed because each closed loop in the Dessin contains
 the point $w=\infty$. These are the new points that we added in order to obtain the Belyi
 pair from the Dessin in section \ref{fromFeynDess}. 
This Dessin is given in Figure \ref{fig:gdessin4}.
 A model for the genus 1 Dessin is more involved, because we need
 both a model for the map and for the torus. 
The fact that there is a single Feynman diagram appearing in the orbit implies  that the Belyi
pair will have all coefficients in $\mQ$. The Belyi curve (worldsheet of 
torus topology) is defined by the 
$$ y^2=x^3-x $$
and the Belyi map is 
$$ w=x^2 $$
The reader can readily verify that, as $ x \rightarrow 0 $, 
we can choose local coordinates $ w = \epsilon_1 $ on the target 
and $ \epsilon_2  $ on the worldsheet 
$ ( y = \epsilon_2 , x = - \epsilon_2^2 ) $
 so that $ \epsilon_1 = \epsilon_2^4 $ as required for a ramification 
point described by a 4-cycle. For $ w=1$, we have worldsheet 
points $ (x=1, y=0) , ( x = -1, y=0)$. Take the first :  
a local coordinate on the worldsheet is $\epsilon_2 $ with   
$ ( x = 1 +  { \epsilon_2^2 \over 2 }  , y = \epsilon_2 ) $. 
On the target a local coordinate is   $( w -1 ) = \epsilon_1$
and locally the map is  $  \epsilon_1 =   \epsilon_2^2$ as required for 
a simple ramification point.  The same argument holds at $(x=-1,y=0) $  
so that we have ramification profile $ [2^2] $ over $w=1$. 
Near $ w = \infty $, we have $ w = x^2 , y^2 = x^{3} $. 
We can choose local coordinate $ \epsilon_1$ on the target 
with  $ w = { 1\over \epsilon_1}  $ and $ \epsilon_2$ on the 
worldsheet with $ ( y = { 1 \over \epsilon_2^3 }  ,  
x = { 1 \over  \epsilon_2^2 )  }  $, 
so that $ \epsilon_1  = \epsilon_2^4 $ as required for 
ramication profile $ [4] $ over $ w = \infty $.

\item  $ \langle tr X^3 tr X \rangle $ : $[3\, 1]\cdot [2^2]=[3\, 1]$\\
 There is a single Hurwitz class and hence a single element in the
 Galois orbit so that again $K=\mQ$. The Belyi map of this Dessin is
 $$ w={(z+{13\over 4})(z+{5\over 4})^3\over z+1} $$
We can easily see that $z = {-13 \over 4 } $ 
has image $ w=0 $ and no ramification, while $ z = { - 5 \over 4 } $ 
also mapping to  $ w=0$  is a third order zero, so 
the ramifcation profile over $w=1$ is $ [31] $.  
Points $ z = { -7 \over 4 } \pm { \sqrt 3 \over 2  } $ 
on the worldsheet map to $ w=1$ and each is 
a second order zero of $ w-1$ thus
 describing ramifcation profile $[2^2]$ over $w=1$. 
The points $ z= -1 $ and $ z = \infty $ map to $ w = \infty $.
While $ z=1$ is a simple pole,  the  large $z,w$ behaviour is $ w \sim z^3 $, 
so that $w=\infty $ has a ramification profile $ [31]$.

\item  $  \langle (tr X^2)^2 \rangle $ : $[2^2]\cdot [2^2]=3[1^4]+2[2^2]$\\
 The contribution coming from $[1^4]$ is disconnected and hence is not considered.
 There is a single Hurwitz class corresponding to $[2^2]$, a single element in the Galois
 orbit and so $K=\mQ$. The Belyi map of the Dessin can be obtained by cleaning from the $\Tr (X^2) $ Belyi map \ref{examp1},
 which gives
 $$ w={4z^2(z-{1\over 4}-z^2)\over (z-{1\over 4})^2} $$

\begin{figure}
\begin{center}
\resizebox{!}{8.5cm}{\includegraphics{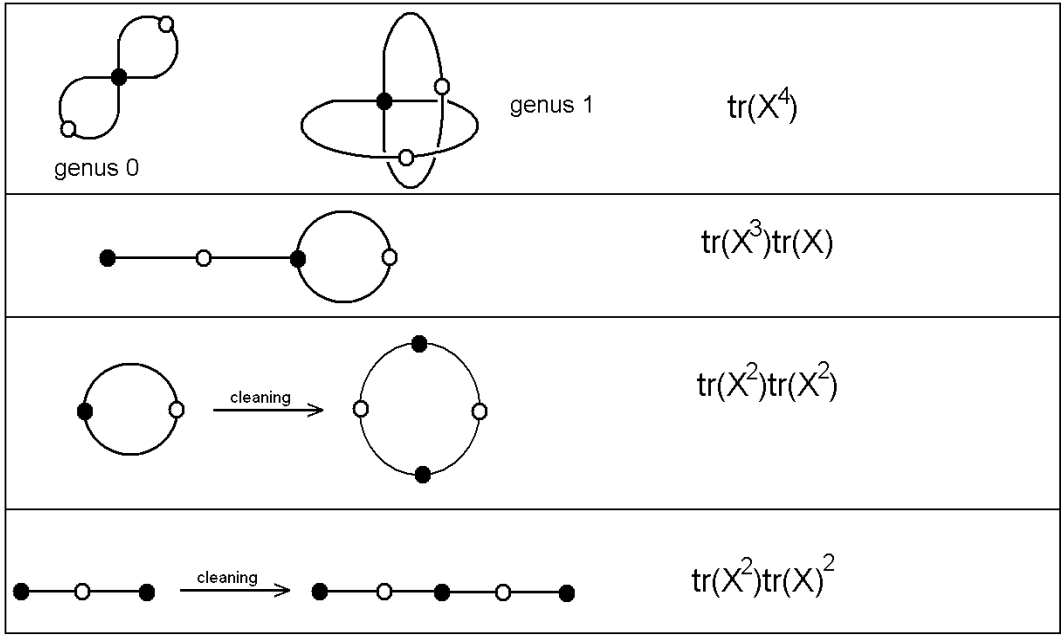}}
\caption{The Dessins corresponding to operators built using four matrices}
\label{fig:gdessin4}
\end{center}
\end{figure}

\item  $  \langle tr X tr X tr X^2 \rangle $ : $[2\, 1^2]\cdot [2^2]=[2\, 1^2] + 2[4]$\\
The contribution corresponding
to $[2\, 1^2]$ is disconnected and hence not considered. There is a single Hurwitz class
corresponding to $[4]$. This class which is related to Chebyshev
polynomials, have a chain Dessin \cite{lanzvon}. There is again a single element in the Galois orbit so that again
$K= \mQ$. The Belyi map of the Dessin can be obtained by cleaning from the
 $ ( \Tr X)^2 $ Belyi map (\ref{examp2}) 
$$ w=16z(1-z)(1-2z)^2 $$

\item  $ \langle (tr X)^4 \rangle$: \\
All contributions are disconnected.

\item
The (single) connected Feynman diagram contributing to the correlator
$$ \left\langle (\Tr X^{k})(\Tr X)^k\right\rangle $$
is given by the Dessin in Figure \ref{fig:gendessin}.
\begin{figure}
\begin{center}
 \resizebox{!}{4cm}{\includegraphics{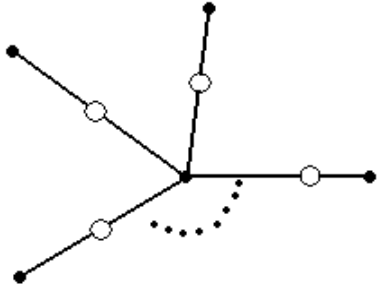}}
\caption{A Dessin for the connected piece of $\langle (\Tr X^{k})(\Tr X)^k \rangle$.}
 \label{fig:gendessin}
\end{center}
\end{figure}
The Belyi function for this flower is\cite{Komatsu}
$$
w = (1+x^k)^2
$$
The field of definition is equal to the moduli field is equal to $\mQ$. Thus, these belong to
a Galois orbit of length 1. Further, 
this Dessin is the only Dessin that arises in the connected
correlator so that this connected correlator gets
 its complete contribution from a single Galois orbit.
By applying the cleaning map to these we would get Dessins that give 
one of many contribution to correlators
of the form
$$ 
(\Tr X^{k})(\Tr X^2)^k (\Tr X)^k 
$$
Thus, for $k>2$ all of these correlators we get contributions from more than one Galois orbit.
Consider the case $k=3$. We have $ [ \sigma_0] = [32^31^3] , \sigma_1 = [2^6] $.
 There are a total of four connected diagrams associated with the 
conjugacy class made of one cycle of length $12$. 
One of  them  is defined over rationals as above 
and lies alone in a Galois orbit.   The other three   lie in
a single orbit; they correspond to the cleaned versions
 of the Dessins shown in Figure \ref{fig:gdessin5}. 
The
 Belyi functions
for the Dessins of Figure \ref{fig:gdessin5} take the form
$$
w=z^3 (z-a_1)(z-a_2)^2
$$
To fix $a_1$ and $a_2$ following  the method of \cite{lanzvon} 
 one needs to pick a root of
\be
25\alpha^3 -12\alpha -24\alpha -16 =0
\label{fod}
\ee
Each root corresponds to a particular Dessin. Given $\alpha$ compute
$$
b={5+4\alpha -\sqrt{(5+4\alpha)^2-62\alpha}\over 12}
$$
and (the six possible roots for $a_1$ all give rotated versions of the single Dessin)
$$
a_1=\left({1\over b^3(b-1)(b-\alpha)}\right)^{1\over 6}
$$
$$ 
a_2=\alpha a_1
$$
The field generated by the defining polynomial (\ref{fod}) has discriminant $-5038848=-2^8 3^9$\cite{PARI}.
This is not a perfect square which tells
us that the Galois group of this field is $S_3$\cite{Milne}.

\begin{figure}
\begin{center}
\resizebox{!}{2.5cm}{\includegraphics{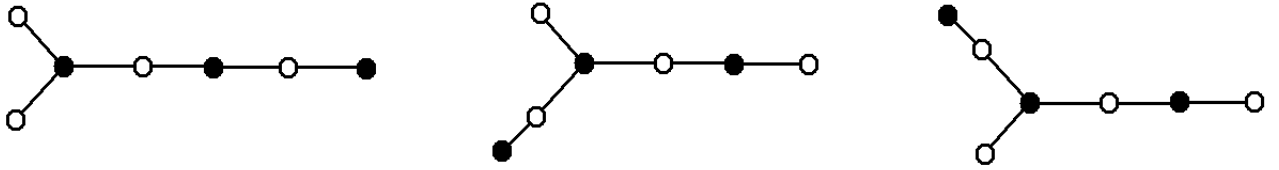}}
\caption{The Dessins shown belong to a single Galois orbit.}
\label{fig:gdessin5}
\end{center}
\end{figure}

\item
There is another interesting example of an operator 
which receives contributions from Dessins belonging to
more than one orbit. The Leila Flowers are a pair of genus zero
Dessins (in fact trees) which can be described by permutations 
\bea
&&  [ \sigma_0] = (1)^{15} (5) \cr
&&   [\sigma_1] =  (2) (3) (4)(5) (6) \cr
&&     [ \tau ] = (20)
\eea
By cleaning (See Figure   \ref{fig:opsforLeilaflower}) 
we have genus zero Dessins with
\bea
&&  [ \Sigma_0 ]  = (1)^{15} (2)(3) (4) (5)^2 (6) \cr
&& [ \Sigma_1 ] = (2)^{20} \cr
&& [ \tau ] = (40)
\eea
Hence there are  at least 2 distinct
Galois orbits contributing to the connected piece of
\bea
\langle   ( tr (X) )^{15} tr (X^2) tr ( X^3)
 tr ( X^4)   (  tr ( X^5) )^2 tr (X^6 ) \rangle
\eea

\end{itemize}

\subsection{  Belyi pairs  and square Strebel differentials }\label{sec:dessribb}

 Zapponi \cite{zappmsri,zapp2000} describes
 how to go from a Dessin to a Ribbon-graph
 which is the critical graph of a
 quadratic differential $\omega$.
 The Dessin is the inverse image of $ [ 0 , 1 ] $.
 The critical graph of the quadratic differential
 is the inverse image of the unit circle. For a quadratic
differential to correspond to a Belyi-pair,
its ribbon graph must be bi-colorable. This is equivalent to
requiring that the quadratic differential is a square of a holomorphic differential.
From the Belyi data the holomorphic differential is $ { d \beta \over \beta } $.
The vertices of the critical graph correspond to the
ramification points over $1$. For clean Belyi maps, these
$d \beta $ have a zero of order $1$, hence $ \omega $ has a zero
of order $2$  and consequently the critical graph has valency $4$.
For the curve defined over $ \bmQ $,
the lengths are all $1$ \cite{mulasepenkava}.
These facts constrain the places
of $ \cM_{ g , n } $ which correspond to clean Belyi maps of degree $2n$.
Since the quadratic differentials and their critical graphs
give a cell decomposition of $ \cM_{g,n} $ as well as a combinatoric
description of Mumford Morita-classes \cite{Kontsevich}, these facts should be useful in developing a
combinatoric derivation of  ELSV type formulae \cite{elsv} in terms
of Mumford Morita-classes for the Hurwitz numbers we have derived.

It is interesting that Strebel differentials, 
and the Galois action on Dessins have been studied in 
a physics context before in connection with Seiberg-Witten theory
\cite{ACD}. Strebel differentials also a prominent role in the 
programme of \cite{gopak}.

\subsection{ Topological string theory on $ \mP^1$ }

We have seen that the 1-matrix model has an interpretation
as a string theory with a $\mP^1$ target space. It is then natural
to look for a topological string theory on $ \mP^1$ which
reproduces the correlators of the matrix model.
A connection between topological $\sigma$-models and
large $N$-matrix integrals has been established in \cite{eguchi}.
This work provides an explicit matrix model which reproduces
the topological $\sigma$-model
(A-model) coupled to gravity on $\mP^1$.
The observables of the topological $\sigma$-model come from the
de Rahm classes of the target manifold. For $\mP^1$ there are
two de Rahm classes, the identity and the K\"ahler class. Denote
the corresponding physical observables $P$ and $Q$. Further,
because the topological $\sigma$-model is coupled to gravity,
new observables given by gravitational descendents of $P$ and
$Q$ can be constructed. Any of these observables can be added
as a term in the action with a coupling $t_P$ ($t_Q$) for
$P$ ($Q$), and with a coupling $t_{n,P}$ ($t_{n,Q}$) for
the $n$th descendant of $P$ ($Q$) respectively. The corresponding
matrix model is
$$ Z=\int dM e^{-{\rm Tr} V(M)} $$
where
$$
\Tr V(M) = -2\Tr M(\log M-1) +\sum_{n=1}2t_{n,P}\Tr M^n(\log M -c_n)+\sum_{n=1}{1\over n}t_{n-1,Q} \Tr M^n
$$
$$
c_0=0\qquad c_n=\sum_{j=1}^n{1\over j}
$$
The matrix $M$ is an $N^*\times N^*$ matrix with $N^*=Nt_{0,P}$.
To get the Gaussian matrix model we should choose $t_{1,P}=1$, $t_{1,Q}=1$ and set all other
couplings equal to zero. Since the perturbation is 
being chosen to cancel the logarithmic term,  
this may well be a subtle limit of the model.

One could also consider a more direct route to constructing
the relevant topological string theory on $ \mP^1$.
It is shown in \cite{zappmsri} how to obtain
Strebel graphs from Dessins. These Strebel graphs
are not the most general. Rather they have the property that
they are related to quadratic differentials which are squares of
ordinary holomorphic differentials. In fact, in terms of the Belyi map,
we have 
$ \omega_{zz} = { \d_z f \over f }{ \d_z f \over f }  $.
Topological string constructions can be designed to
localize on the solution sets of appropriate
equations. This equation should form the basis of a topological
string construction with $\mP^1$ target where the
constraint of three branch points is automatically
included.

Some insights on the string dual of 2dYM, which 
has a string theory interpretation in terms of 
two-dimensional target spaces $ \Sigma_G$ ,  have been 
obtained by developing an interpretation in terms 
of a six-dimensional Calabi-Yau target space which are bundles 
over $ \Sigma_G$\cite{2dYM6D}. It would be interesting in the 
case of the string dual of the Hermitian Matrix model to explore if 
a six-dimensional target space interpretation can 
provide a natural home for the $\mP^1$ with three 
branch points discussed here.

\section{ Multi-Matrix models and colored-edge Dessins  }\label{MMCED}

\subsection{ Multi-Matrix models }

Consider  the Gaussian 2-matrix model and a multi-matrix operator 
$tr ( \sigma X^{\otimes 2n_1} \otimes Y^{ \otimes 2n_2 })$. The correlator is given by 
\bea
\langle tr ( \sigma X^{\otimes 2n_1} \otimes Y^{ \otimes 2n_2 } )  \rangle
&& = \sum_{ \g_1 \in [2^{n_1} ] } \sum_{ \g_2 \in [2^{n_2} ] }
 tr_{ 2n } ( \sigma \gamma_1 \circ \gamma_2 ) \cr
&& =  \sum_{ \tau \in S_{2n} }  \sum_{ \g_1 , \g_2  }
    \delta ( \sigma  ( \g_1 \circ \g_2 ) \tau ) N^{ C_{ \tau }}
\eea

The choice of an operator, say $ tr X Y tr XY $,
chooses a coloring of the cyclically ordered edges coming out of a black
vertex (associated with $ \s_0 $ in Belyi literature conventions, here
$ \s $).
The choice of a contraction gives a clean Dessin,
equipped with the additional data of a coloring of
the edges. Any open circle (associated with $\s_1$ in Belyi literature conventions, here $ \s_1 = \g_1 \circ \g_2 $) has 2-edges of the same color,
which form the propagator in the matrix theory language.
We can think of $X$ and $Y$ propagators as having different colors.
For an earlier use of related ideas see \cite{Kazakov:1986hu}.

In the case of the 1-matrix model, we normalized the
1-point functions by introducing $ { | [\s] | \over (2n)! } $.
Here $ 2n_1 + 2n_2 = 2n $  and the natural normalization factor is
\bea
{ | [\s ]_{ ( 2n_1, 2n_2 )  } | \over ( 2n_1)! ( 2n_2)! }
\eea
where
$  [\s ]_{ ( 2n_1, 2n_2 )  } $ is the set of permutations in $S_{2n}$
related to $ \s $ by conjugation with
 elements in $ S_{2n_1} \times S_{2n_2} $.   This is analogous to
the fact that $ [ \s]$ was the set of elements related to $ \s $
by conjugation in $S_{2n}$,  equivalently the elements in the
same conjugacy class as $ \s $ or elements with same cycle structure.
A natural guess now is that the answer will be related to
$ Aut  ( D_{col} ) $, the automorphism group of the colored
Dessin. It is clear that the colored Dessin  $D_{col} $  gives
a  clean Dessin $D$ of usual type (familiar from Belyi literature)
by forgetting the colors.
This has $ Aut ( D ) $ which is the same as the $ Aut ( f  ) $,
 the automorphism
of  the holomorphic map. As a step towards the definition of
 $ Aut ( D_{col} ) $, note that two colored Dessins
 are equivalent when they are related
by permutations in $ S_{2n_1} \times S_{2n_2} $.
We are lead to
\bea
Aut ( D_{ col} ) = Aut ( D ) \cap   ( S_{2n_1} \times S_{2n_2} )
\eea
Both $ Aut ( D ) $ and $   ( S_{2n_1} \times S_{2n_2} ) $ are subgroups
of $ S_{2n} $ and the intersection is a subgroup.
This symmetry appears in the appropriately normalized correlator
\bea\label{corautcol}
 { | [\s ]_{ ( 2n_1, 2n_2 )  } | \over ( 2n_1)! ( 2n_2)! } N^{ C_{\s} - n }
   \langle tr ( \sigma X^{\otimes 2n_1} \otimes Y^{ \otimes 2n_2 } )  \rangle
&& = \sum_{ \s \in [ \s]_{2n_1 , 2n_2} } \sum_{ \tau , \g_1 , \g_2 }
 { N^{ \chi ( D ) } \over ( 2n_1 )! ( 2n_2)! }
 \delta ( \sigma  ( \g_1 \otimes \g_2 ) \tau ) 
 \cr
&&
= \sum_{ D_{ col } } { N^{ \chi ( D ) }  \over | Aut ( D_{ col} ) |  }
\eea
The last line follows from a general fact about group actions.
In  this case there is an action by conjugation
 of $S_{2n_1} \times S_{2n_2}$ on  the permutations solving the delta function
and  the multiplicity of equivalence classes is given by the
order of the cosets
$ { ( 2n_1 )! ( 2n_2)! \over | Aut ( D_{ col} ) | }$

The remark of section  \ref{cominteg}  generalizes to the colored case.
\bea
 N^{ C_{\s} - n }
 \langle tr ( \sigma X^{\otimes 2n_1} \otimes Y^{ \otimes 2n_2 } ) \rangle
= \sum_{ [ D_{col} ( \sigma  ) ] }
{ | Aut_{2n_1, 2n_2}  (  \sigma ) |  \over  | Aut ( D_{ col} ) | } N^{ \chi ( D ) }
\eea
The $ Aut_{ 2n_1, 2 n_2} ( \sigma ) $ is the group  of
permutations in $S_{2n_1} \times S_{2n_2} $ which leaves invariant
$ \sigma \in S_{ 2n } $
\bea
| Aut_{2n_1, 2n_2}  (  \sigma ) | = { ( 2n_1) ! ( 2n_2) ! \over
 |[ \sigma ]_{2n_1, 2n_2} |   }
\eea
The ratio $ { | Aut_{2n_1, 2n_2}  (  \sigma ) |  \over  | Aut ( D_{ col} ) | } $
is integral because $  Aut ( D_{ col} ) $ is the intersection
in $ S_{2n_1} \times S_{2n_2} $ of $ Aut_{ 2n_1 , 2n_2}  ( \s ) $
and $ \gamma_1\circ \gamma_2  \in ( [2^{n_1} ] , [ 2^{n_2} ] ) $. Hence it is a subgroup
of  $ Aut_{ 2n_1 , 2n_2}  ( \s ) $. The ratio is the number
of Wick contractions of $ tr ( \sigma ( X \otimes Y ) ) $
which give the colored-Dessin-equivalence class $D_{ col}$.
As before we also deduce the integrality of
\bea
{  2^{n_1 + n_2 } n_1! n_2 !  \over Sym ( [ \s ]_{ 2n_1 , 2n_2}   ) }
 \langle tr ( \sigma X \otimes Y )  \rangle
\eea

Note that { \it the set }  of $ D_{ col} $ for a given $ D$ is
clearly a property of the Hurwitz-class, i.e of the conjugacy
class of $ ( \s_0 , \s_1 ) $ under simultaneous conjugation
by $ S_{2n} $, hence its a property of an equivalence 
class of holomorphic maps.
This remark will be exploited in section \ref{CEDGI} to build Galois
invariants  of the Hurwitz class.

\subsection{ Colored Dessins,  permutation triples and  subgroups of $ S_{2n}$  }
\label{sec:colpersub} 

In the classic case we have the correspondence between

\begin{itemize}

\item Triples of permutations $ \sigma_0 , \sigma_1 , \sigma_{\infty}  $
in $S_{2n} $ obeying $ \s_0 \s_1 \s_{\infty}  =1 $, up to
equivalence under simultaneous
conjugation in $S_{2n}$. Further permutation the $\s_1$ belongs to the
conjugacy class $[ \s_1 ]  =  [2^{n} ] $.

\item Clean Dessins d'Enfants which are graphs with alternating 
black and white vertices. The cyclic order for the edges at each 
vertex is part of the data of the graph.
The white vertices have valency $2$.

\end{itemize}
This correspondence implies that the automorphism of a triple 
is a symmetry of the Dessin.
There is more we could add to this list of remarkable
equivalences, including Hurwitz classes and Belyi-pairs of $  ( X , \beta ) $
defined over $\bmQ$.  To motivate our generalization, the
above list is rich enough. 

The above list generalizes to colored Dessins as follows

\begin{itemize}

\item Triples of permutations $ \sigma_0 , \sigma_1 , \sigma_{\infty}  $
with $ \s_0 \in S_{2n}$,  $ [\s_1 ] \in [ 2^{n_1} , 2^{n_2} ] $ in
an $ S_{2n_1} \times  S_{2n_2}$
subgroup of $ S_{2n} $  obeying $ \s_0 \s_1 \s_2 =1 $ up to
equivalence under simultaneous conjugation in
$ S_{2n_1} \times S_{2n_2} $.

\item Clean Dessins d'Enfants which are again graphs with alternating black and white vertices.
The white vertices again have valency $2$. In the colored Dessin, the edges can be red or blue.
The edges incident on a white vertex are both red or both blue. The edges ending on a black
vertex can have any color.

\end{itemize}

Automorphisms of the colored Dessin defined as conjugations in
$ S_{2n_1} \times S_{2n_2} $ which leave the triple fixed, coincide
with the symmetry group of the colored Dessins.

\begin{figure}
\begin{center}
 \resizebox{!}{4cm}{\includegraphics{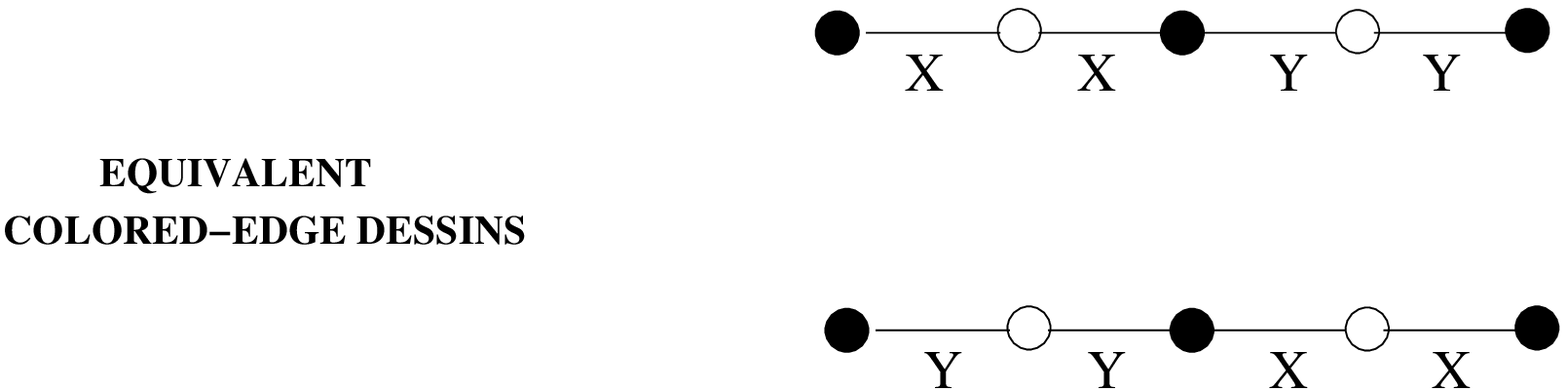}}
\caption{Colored Dessins in the same equivalence class }
 \label{fig:equivCDEs}
\end{center}
\end{figure}

We have given formal definitions of equivalence 
of the colored Dessins. In simple examples, 
this reduces to fairly obvious-looking equivalences. 
Two simple colored Dessins which are in the same equivalence class, 
 are shown in 
Figure \ref{fig:equivCDEs}. 

The following question is natural : What continuum
object does a choice of  $ D_{ col} $ correspond to ?
We will develop two related answers to this question.
One is that a choice of $D_{col } $ is a pair consisting
of $ ( f , s)  $, a holomorphic map from a Riemann surface to
$\mP^1$ with branch points at $ 0 , 1 , \infty $
and a section $s$  of a skyscraper sheaf on the
Riemann surface. The second answer is that the colored Dessins
define  sheaves over Hurwitz space.

\subsection{ Coloring edges to coloring vertices }
The Dessins that arise as Feynman diagrams of
the Hermitian matrix model are {\it clean}  Dessins.
Each propagator is a pair of edges which are joined
at a white vertex. These white vertices are inverse images of
$1$. Coloring the propagators is equivalent to coloring the
white vertices instead and leaving the edges uncolored.
Each Dessin determines a Riemann surface and
an equivalence class of maps from the Riemann surface
to $\mP^1$ with three fixed branch points, chosen to
be at $0 , 1 , \infty $. Recall that the white vertices correspond
to inverse images of $1$, which are ramification points of order $2$.
Each colored Dessin associates colors to these ramification points.
This can be described in terms of skyscraper sheaves on the worldsheet, 
localised at the ramification points above 1, which associate
the set of colors to these points.(We have found \cite{koppe} 
to be a useful reference on the basics of sheaves we need for this discussion). 
 Choosing a coloring is a choice of
sections of the skyscraper sheaves. The Automorphism group
$ Aut D_{col} $ can then be equated to $ Aut ( f , s ) $
which is the automorphism of the pair consisting of the map from
the worldsheet with simple ramification points above $1$, along with a
coloring of the ramification points. For the Dessins corresponding
to the map $f$, the automorphisms are just maps $ \phi $
obeying $ f \circ \phi = f $. Such maps $ \phi$
have to map the set of ramification points above $1$ back to itself.
To define $ Aut ( f , s ) $  we also require that $ \phi $
preserves the color at the ramification points.  So we can say that
\bea
 { | [\s ]_{ ( 2n_1, 2n_2 )  } | \over ( 2n_1)! ( 2n_2)! } N^{ C_{\s} - n }
   \langle tr ( \sigma X^{\otimes 2n_1} \otimes Y^{ \otimes 2n_2 } ) \rangle
 = \sum_{ D_{ col } } { N^{ \chi ( D ) }  \over | Aut ( D_{ col} ) |  }  =   \sum_{ ( f , s )  } { N^{ \chi ( D ) } \over | Aut ( f , s  ) |  }
 \eea

\subsection{  Sheaves of  colored Dessins over Hurwitz space }\label{shcodes}

Consider the  Hurwitz space with 3 branch points
at $ 0, 1, \infty $ on $\mP^1$. Consider small non-overlapping open discs
drawn on the $\mP^1$. If we let  the branch points move over
their respective discs, there is a 3-complex dimensional subspace of
Hurwitz space. The $S_d$ equivalence classes of
triples $ [ \sigma_0 , \sigma_1 , \sigma_{\infty}  ] $ describe different
strata of Hurwitz space. Each of these strata is associated to a
Dessin corresponding to  $ [ \sigma_0 , \sigma_1 , \sigma_{\infty}  ] $.
For our current applications we are interested in strata
where $ d = 2n $ and  $ [\sigma_1] = [ 2^n ] $.
For each choice of a positive integer $ k$, we can consider
colorings of the Dessins with $k$-colors, where we can
further  specify a partition of $n$ of length $k$, i.e
$ n = n_1 + n_2 + \cdots + n_k $. For each $ (  k , \vec  n = ( n_1 , \cdots , n_k ) )$, 
we have a set of equivalence classes of colored Dessins.
Equivalently, as explained above we have the pairs $ ( f , s ) $.
We can define  a sheaf over the set of Grothendieck  Dessins by associating
to each Dessin the set of equivalence classes of colorings of that  Dessin.
Since strata of Hurwitz space map to Dessins, we can pull back
the sheaf of colored Dessins to the open three-dimensional
regions of Hurwitz space described above.  This gives one answer to
the question of the continuum interpretation of the colored Dessins.
They are related to sheaves over Hurwitz space.
We leave a more detailed and general discussion of such sheaves
to the future, including the questions of how to extend
the definition to compactified Hurwitz spaces where branch points
are allowed to collide, Dessins degenerate and non-trivial restriction
maps of  sets of colored Dessins arise.

\subsection{ Hurwitz space and String theory for multi-matrix models }

In  section  \ref{onemathurwitz} we developed  an interpretation of
the one-matrix model at generic couplings
in terms of a topological  string theory which localizes on
 the Hurwitz space of holomorphic maps  with three branch points
on $ \mP^1$ target space. 

In this section we have shown that the Hurwitz spaces for
$ \mP^1$  target and 3 branch points continue to
provide a string theory interpretation for multi-matrix models
at generic couplings. Now it is a string theory
related to sheaves over Hurwitz space.

It would be interesting to develop a more physical
approach to this string theory, e.g as topological
sigma model coupled to 2D gravity, or as some topological
WZW CFT coupled to 2D gravity, perhaps along the lines of
\cite{wittenN} where a physical set-up for
intersection theory for bundles over $  \cM_{g,n}$
is described.

\section{ Colored-edge Dessins as a tool for  Galois invariants  }
\label{CEDGI}

In the following we will refer to Grothendieck Dessins
(ribbon graphs) as Dessins for short. We will use the 
abbreviations CEDs for colored-edge Dessins 
and MMOs for multi-matrix operators.\footnote{We apologize 
to  readers who hate acronyms and
appeal to their concern for the trees we save. }

\subsection{ From Dessin to  list of colored-Dessins and  list of multi-matrix operators }
\label{sec:desstolists} 

\begin{figure}
\begin{center}
 \resizebox{!}{7cm}{\includegraphics{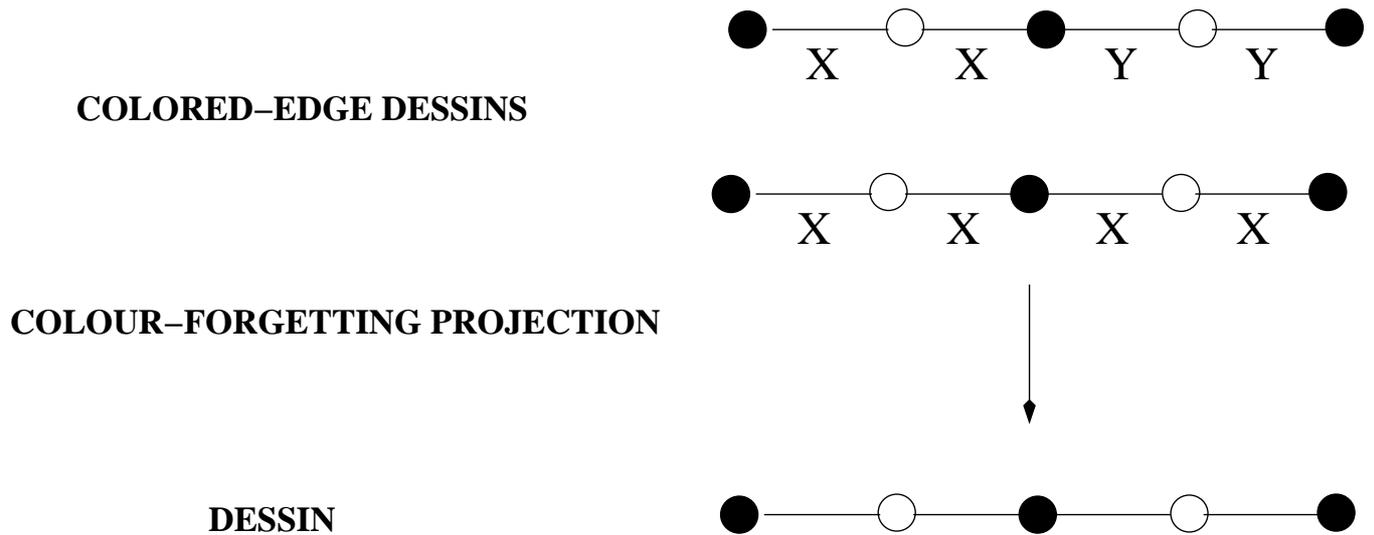}}
\caption{Equivalence classes of colored Dessins and projection to Dessin }
 \label{fig:simp-col-dess}
\end{center}
\end{figure}

We showed in section  \ref{MMCED} that
the 1-point function of a  multi-trace operator
in multi-matrix theory is obtained by summing 
over  certain triples of permutations which 
correspond to colored-edge Dessins (CEDs).  These CEDs
project to ordinary Dessins with black and white vertices
by forgetting the edge-colorings.
A simple example illustrating the concept 
of  color-forgetting projection map from the set of 
equivalence classes of colored Dessins to ordinary Dessins
is given in Figure \ref{fig:simp-col-dess}. 

 We have labeled the colors as $X , Y $. 
 Since the correlator of a
 given multi-matrix operator is a sum over colored-edge
 Dessins,  multiple CEDs are associated to a single
 multi-matrix operator (MMO). Forgetting the coloring 
 gives an ordinary Dessin. 
 We will show that the set of Dessins  associated to a 
 multi-matrix operator in this way do not form
 complete Galois orbits, unlike the case of the one-matrix model.

One can also associate   a list of MMOs with
a  Dessin. For a fixed  Dessin there is  list of MMOs 
receiving contributions from the colorings of that Dessin, in other 
words, from  the set of colored Dessins which project to the given Dessin.

The two-color-Dessin in Figure \ref{fig:simp-col-dess}
 arises when we consider the 
one-point function of $ tr X tr Y tr XY $ in the 
Gaussian 2-matrix model. The fact that one black vertex 
has a single  incident edge labeled with a single $X$, 
another black vertex has a single incident edge labeled with a single $Y$, 
and the middle black vertex has two incident edges 
labeled $X , Y $, reflect the structure of the operator. 
Colored Dessins, like ordinary Dessins, come equipped 
with a cyclic order at each vertex. This corresponds to the 
cyclic property of traces. The one-color Dessin 
in Figure \ref{fig:simp-col-dess} arises from considering 
the correlator of  $ (tr X )^2 ( tr X^2 ) $ in the Gaussian 
1-matrix model. So we can associate, to the uncolored Dessin 
in the figure, two distinct MMOs, namely  $ (tr X )^2 ( tr X^2 ) $
and $  tr X tr Y tr XY $.  We will explain shortly 
how these lists of colored Dessins or lists of MMOs 
provide combinatoric Galois invariants.

A general Dessin where the white vertices 
do not all have exactly  two incident edges, can as discussed 
in section (\ref{3tobarQ}) be converted  into a clean
 Dessin by turning the white vertices 
into black vertices, and introducing white vertices 
in the middle of the existing edges. 
In the Galois 
theory literature, one performs this operation to 
define the { \it cartographic group} of a Dessin, which 
is a Galois invariant. We can also use this procedure of 
cleaning to associate 
to a Dessin combinatoric Galois invariants related 
to lists of CEDs or lists of MMOs.

\begin{figure}
\begin{center}
 \resizebox{!}{7cm}{\includegraphics{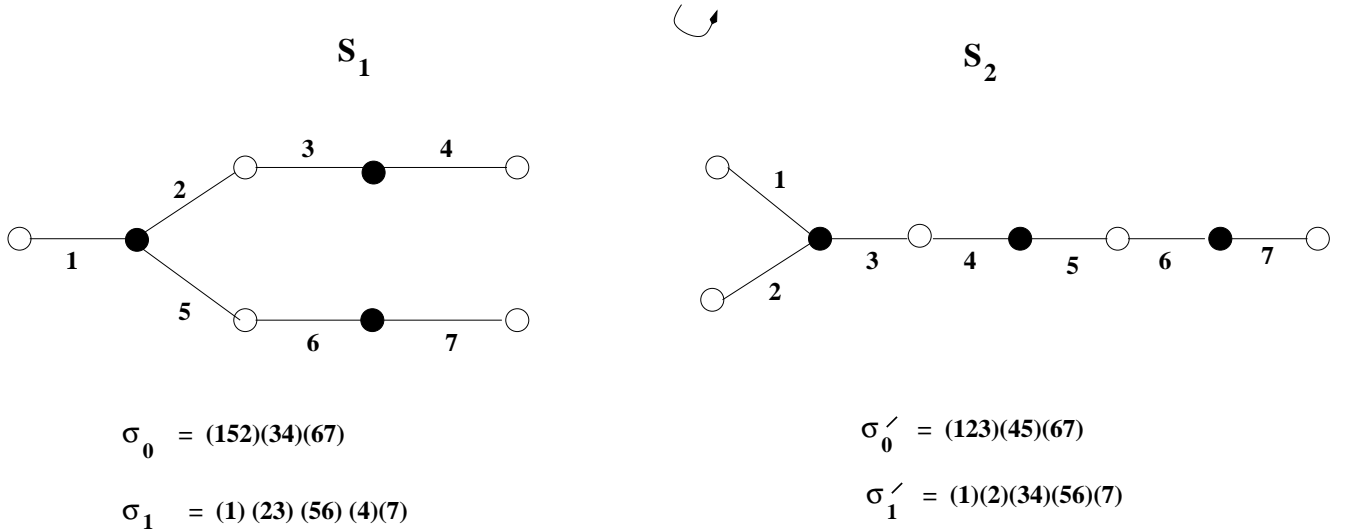}}
\caption{Two distinct Dessins with same cycle structure for permutations }
 \label{fig:S1S2}
\end{center}
\end{figure}

\begin{figure}
\begin{center}
 \resizebox{!}{7cm}{\includegraphics{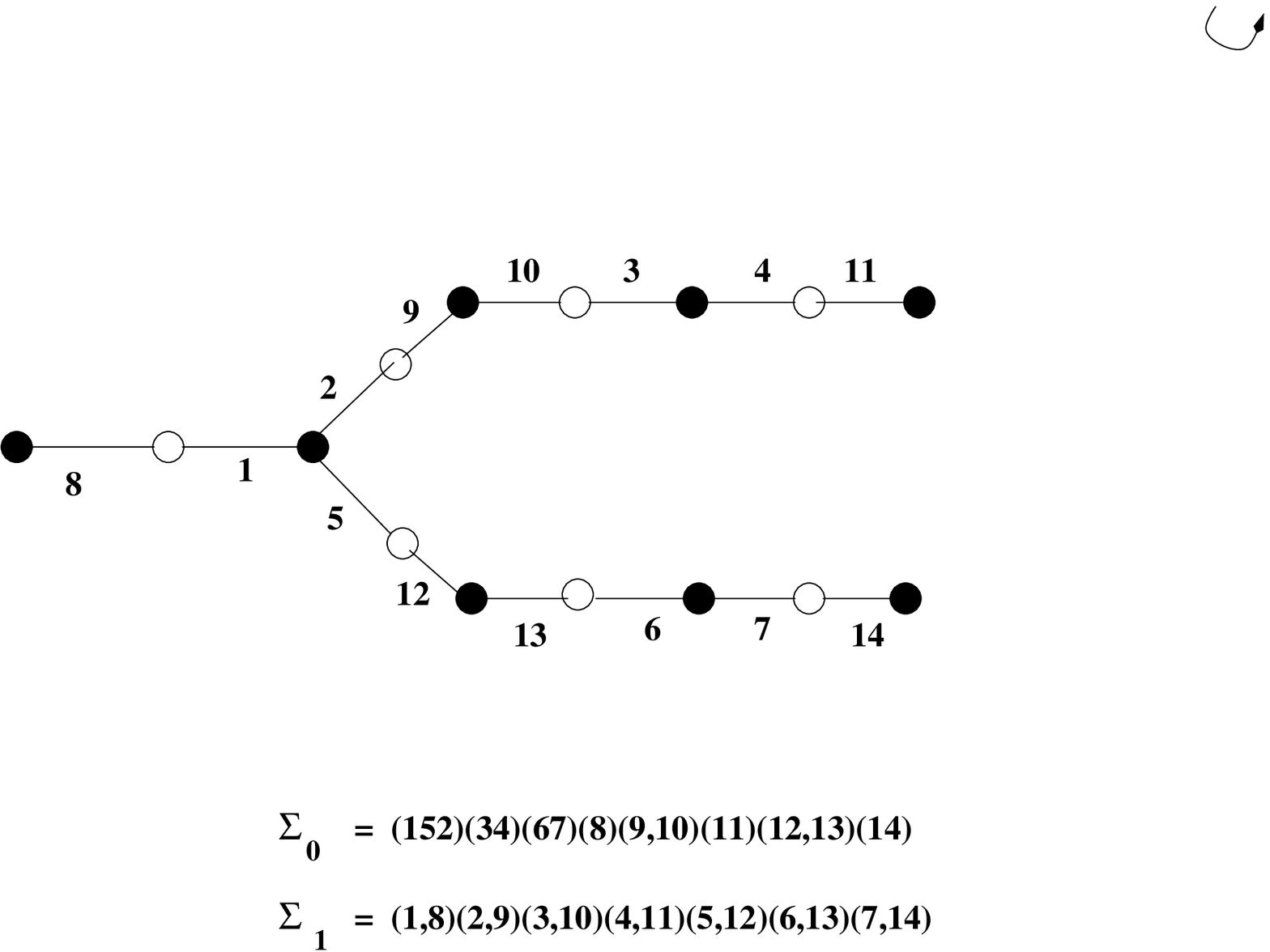}}
\caption{ Cleaned version of S1 and permutations }
 \label{fig:cleanedS1}
\end{center}
\end{figure}

We will describe a 
convenient way to generate lists of possible 
MMOs for a given Dessin with a given number of 
different matrix types, which can be done with a 
computer program \footnote{code written in  SAGE available from 
authors if desired}.  For concreteness, we will describe 
the construction with the aid of an example. 
Take two Dessins as in Figure \ref{fig:S1S2}
which can be described by permutations as in the Figure.   
For the Dessin $S_1$ we have two permutations $ \s_0 , \s_1$
which can be read off by going round the black and white vertices 
respectively, in the orientation indicated. The 
same procedure produces a pair of  permutations 
$\s_0^{\prime}  , \s_1^{\prime} $ for the Dessin $S_2$.
The cycle structures of the $ (\s_0 , \s_1)$ are the 
same as $ (\s_0^{\prime} , \s_1^{\prime} )  $ but 
there is no  permutation which can simultaneously 
conjugate the first pair to the second. So these are 
inequivalent Dessins. They are not clean Dessins. 
After applying the cleaning procedure, the cleaned 
version of $S_1$ is described by permutations 
\bea 
&& \Sigma_0  = (152) (34) (67) (8) ( 9, 10 ) (11)   
(12,13) (14) \cr 
&& \Sigma_1 = ( 1, 8 ) (2,9 ) (3,10) ( 4,11 ) (5,12) (6,13) (7,14) 
\eea 
The permutation $ \Sigma_0 $ is written down by composing 
$ \sigma_0$ with a shifted version of $ \sigma_1$ (see Figure 
\ref{fig:cleanedS1}).
The cleaned version of $ S_2$ is described by permutations 
\bea 
&& \Sigma_0^{\prime} = (123)( 45) (67) (8) (9) (10,11) (12,13) (14)   \cr 
&& \Sigma_1^{\prime} = ( 1, 8 ) (2,9 ) (3,10) ( 4,11 ) (5,12) (6,13) 
(7,14)                                
\eea  
To list the MMOs associated with these clean Dessins, 
proceed as follows. Fix a number of matrix types, 
say  for example $2$, i.e we are looking at $ X , Y $. 
Fix the number of each matrix type, say $ 2n_1 = 10 $ $X$'s
and $2n_2 = 4$ $Y$'s. There are 
$ { n! \over n_1! n_2! } = { 7! \over 5! 2! }  $
ways of distributing $ X, Y $ among  $1, \cdots 7 $. 
Each of these choices leads to a multi-matrix operator. 
By scanning this list we can get all the possible 
MMOs for two matrix types whose correlators in the 2-matrix model
 receive contributions from colored versions of each Dessin.   

It is known that $S_0$ and $S_1$ are in the same Galois orbit
(see page 90-91 of \cite{lanzvon} and \cite{wiki:dess})  
In the above case, we see that the operator $ (Y^2X) (YX)^2 (X^2)^2 (X)^3 $ 
appears in the list for the cleaned version of $S_1$ 
but not in the cleaned version of   $S_2$. Its appearance 
in the cleaned version of $S_1$ arises by associating colors as  
$ (2,9) \rightarrow Y , ( 5,12) \rightarrow Y $ and  
remaining pairs of edges in cleaned $S_1$ as  $X$.  

This example illustrates an important point,
 which was a priori not obvious to us.
 In the case of the Gaussian 1-matrix model with matrix $X$, 
an operator with non-zero 1-point function has $2n$ copies of $X$. 
The different ways of tracing are described by  permutations
$\sigma_0 \in S_{2n}  $. Different elements in the conjugacy class 
$ [ \sigma_0  ] $ give the same operator. The computation of 
its correlator involves a sum over all permutations $ \sigma_1 $
in the conjugacy class with $n$ cycles of length $2$. 
This sum includes all possible Dessins with the specified 
conjugacy classes of $ [\sigma_0] , [\sigma_1]$.
Two Dessin in the same Galois orbit necessarily have the 
same $ [ \sigma_0 ] , [ \sigma_1] $. This means that 
correlators in the 1-matrix model sum over complete Galois orbits. 
In the case at hand, we have seen two Dessins in the 
same Galois orbit, one of which has the property that its 
colorings include one which contributes to the 
1-point function of  $ (Y^2X) (YX)^2 (X^2)^2 (X)^3 $ 
whereas the other Dessin does not have a coloring 
which contributes to the 1-point function of 
 $ (Y^2X) (YX)^2 (X^2)^2 (X)^3 $. So we conclude that in general 
multi-matrix operator correlators do not receive 
contributions from complete Galois orbits of Dessins.

\subsection{ Some new combinatoric Galois invariants }

We  fix a number $k$ of Matrix types.
We specify a vector of positive integers
$ \vec n = ( n_1 , n_2 \cdots , n_k )$ of
length $k$,  which determines how many matrices  of
each type we have. For each Dessin, 
we get a list of multi-matrix operators.  
If we take the intersection of the lists 
associated with each Dessin  in an orbit 
we get a Galois invariant. In examples where we know a set of Dessins 
to form a complete Galois orbit, we 
can compute the lists for each Dessin
and then determine the intersection. 
To make this a powerful idea in the 
determination of Galois orbits we would need to 
find some method, which, just by considering the 
list of MMOs for a single Dessin can determine, 
without prior knowledge of the orbit structures, 
which of the MMOs belong to the intersection.

Similarly we can consider the union 
of lists over a Galois orbit, which gives another 
Galois invariant. The intersection appears a
more economical invariant to consider.

An analogy  is that the moduli
field of a Dessin $K_D$ is not Galois invariant.
Although it does contain some Galois-invariant
information such as the length of the Galois orbit
which is equal to $ deg ( K_D : \mQ )$.  
But the moduli field of the orbit,
which is the normal extension of $K_D $ and contains
all the moduli fields of the individual Dessins, is
a Galois invariant \cite{lanzvon}. 

A natural question is whether these 
combinatoric multi-matrix Galois invariants are a complete set ?
The union invariant is complete. By choosing $k$ to be as large as the
number of edges in the Dessin, we can determine permutations
$ \sigma_0 , \sigma_1$ which describe the Dessin. 
Since the $S_d$ equivalence class of
  $ \s_0, \s_1$ completely identify the Dessin, 
the maximal number of colors certainly determines 
the Galois orbit which the Dessins sits in. 
By using fewer colors, we may hope to extract information 
that identifies the orbit only and not extra information 
about the Dessin. It is not at all obvious that the intersection invariant 
is complete. 

It is worth remarking that we have considered 
the construction of Galois invariants from 
lists of MMOs, but we could equally well have considered 
lists of CEDs, then taken intersections and  unions.

We leave it to future research to determine
effective ways of computing the intersection 
invariant, or similar invariants from lists of 
colored edge Dessins or multi-matrix operators, 
and to determine their usefulness. What is clear is that 
lists of CEDs and MMOs capture more detailed 
information about Desssins than coarse Galois  
invariants such as $ [\s_0 ] , [ \s_1 ] , [ \s_0 \s_1] $. 
In the following section, we will make use of this observation to 
translate known Galois invariants into the language of CEDs and MMOs.  

\subsection{ Known invariants and edge-colorings of Ribbon graphs }

\subsubsection{ Flower-shaped trees  }

There is a famous case of two trees, which are in different Galois orbits,
but require a rather non-obvious Galois invariant to separate them 
\cite{schneps}. 
Both trees have a central black  vertex, and 5 white vertices joined to it.
These 5 white vertices have, respectively, 1, 2, 3, 4, 5 black  vertices
attached to them. The ordering of these 5 white vertices is different
in the two  trees. According to Zapponi \cite{zappmsri}
the action of the Galois group is to permute the five petals to give
a permutation in $S_5$ defined up to multiplication
by the 5-cycle $(12345)$, which means that the sign of the
 permutation is invariant, so there are two distinct Galois orbits 
among Dessins of this type.

This description of the Galois invariant in terms of the sign
of a permutation of the petals can be mapped to a description
in terms of multi-matrix operators or colored Dessins 
associated to the  specified Dessin.

\begin{figure}
\begin{center}
 \resizebox{!}{7cm}{\includegraphics{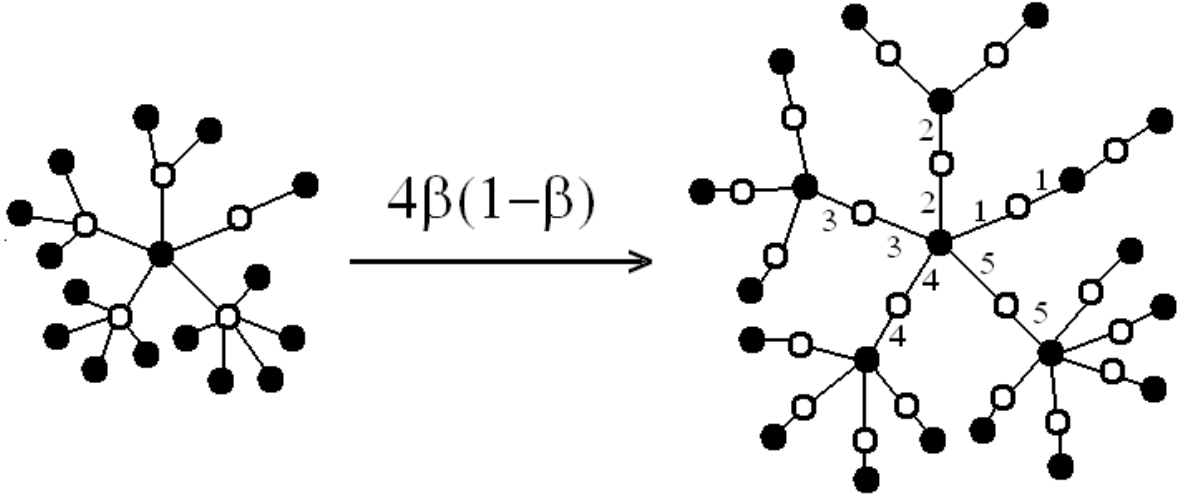}}
\caption{A  multi-matrix operator from Leila Flower }
 \label{fig:opsforLeilaflower}
\end{center}
\end{figure}

We will use $5$ types of matrices  $ Y_1 , Y_2 , Y_3 , Y_4 , Y_5 $
along with the matrix $X$. 
We will focus on operators where the central vertex has these
$5$ $Y_i$. The remaining edges are labeled by $X$. 
In the list associated with the first tree, we will have
\bea
 ( Y_1 Y_2 Y_3 Y_4 Y_5) ( Y_1X ) (Y_2X^2 ) ( Y_3X^3 ) ( Y_4X^4) (Y_5X^5 )
(X)(X^2)(X^3) (X^4)(X^5)
\eea  In the list for the second tree,
the multi-matrix operator will be
\bea
  ( Y_1 Y_2 Y_3 Y_4 Y_5) ( Y_2X ) (Y_1X^2 ) ( Y_3X^3 ) ( Y_4X^4) (Y_5X^5 )
(X)(X^2)(X^3) (X^4)(X^5)
\eea
 So clearly we can express the permutation
relating the two configurations in terms of a permutation
relating operators in the respective lists, constrained to
have $Y$'s at the central vertex. Hence the Galois invariant
can be expressed in terms of combinatoric information 
about the Dessin encoded by the lists of MMOs associated to it. 
In the Figure \ref{fig:opsforLeilaflower},
 we show one of these trees and we also show how the tree changes when
we apply $ \beta \rightarrow 4 \beta ( \beta -1) $,
and the corresponding multi-matrix operator.

Galois orbits are understood for more general classes 
of trees related to the one above. If the valencies  
$(2,3,4,5,6)$ of outer white vertices  are replaced by 
$(a_1,a_2, a_3, a_4, a_5 )$, then we have two orbits 
if $ a_1a_2a_3a_4 a_5(a_1+a_2+a_3+a_4+a_5)$
 is a square and one orbit otherwise.

We observe that the data $(a_1,a_2,a_3,a_4,a_5)$ which is 
important in determining Galois orbits can also be described 
in terms of CEDs. Again we color the central edges 
$Y_1, Y_2 , \cdots Y_5$.  The $X$'s are replaced by $X_1, X_2  \cdots  $
 with the
condition that the $X$'s going with $Y_1$ have lower
indices than those going with $Y_2$ etc. Then we have
$ ( Y_1 X_1)  , ( Y_2 X_2 X_3 ) .. $. Now if we keep the 
colors of edges connected to the central vertex fixed and 
let colorings vary only in  the cycle containing $Y_i$ along with $X$'s
we get the numbers $p_i! $. From this we can extract the $a_i = p_i + 1 $.
The condition on $ a_1a_2 .. a_5 ( a_1 + \cdots + a_5) $
can now be expressed in terms of these multiplicities
of CEDs or of MMOs.

\subsubsection{Galois orbits  for  $K_{n,n}$  }\label{sec:JSWinvts} 

Consider Dessins with underlying graph consisting  
of $n$ black vertices and $n$ white vertices. 
To get the underlying graph from a Dessin, 
just forget the cyclic ordering at each vertex. 
Further restrict to the case where each of the $n$ 
black vertices is connected to each white vertex 
by exactly one edge. This is called a
complete bi-partite graph $K_{n,n}$. 
Further restrict to the case $ n=p^e$ for 
a prime $p$ which is not equal to $2$ and positive integer $e$.  
The set of all Dessins with this underlying graph
are  { \it regular} 
and the structure of their Galois orbits is known \cite{jsw}.
Regular Dessins have an automorphism group which is transitive, i.e 
when viewed as a subgroup of $S_d$ 
can map any integer from $\{ 1 , \cdots , d \} $ to any other. 
There are $p^{e-1}$ distinct Dessins. These are organized 
into orbits parameterized by integers $f$ 
from the set $ \{ 1 , 2, \cdots e \} $. For each
$f$ consider integers $u$ from the set 
$ \{ 1 , 2, \cdots , p^{e-f})$, which are not divisible 
by $p$. There are $ \varphi ( p^{e-f} ) $  such choices,
where $ \varphi $ is the Euler totient function. Each  choice
 leads to a curve $ X_{p,e,f,u} $ and a Belyi map to $\mP^1$.
Using an identity $ \sum_{f } \varphi ( p^{e-f} ) = p^{e-1} $ 
we have the number of Dessins stated above. 

The construction of these Dessins is given 
a group theoretic description in  \cite{jsw}. 
It is useful, for our purposes, to find explicitly 
the description  in terms of permutations $ \sigma_0 , \sigma_1$. 
We know from the description above that the degree of the 
map is $n^2$. Above each of $0$ and $1$, 
there are $n$ ramification points 
each with degree $n$. So we are looking for 
permutations in $S_{n^2}$. 

Define $q= p^f +1$, and consider 
\bea\label{ghn}  
&& h ( nk+ 1+ i ) = nk +1+  (i+1)_n \cr 
&& g^{-1} ( nk +1+ i ) = n ( k+1 )_n + 1 + ( q  i )_n \cr 
&&  g  ( nk + i +1 ) =  n ( k-1)_n + 1 + ({i\over q})_n
\eea 
with $ i = 0 \cdots n-1 $ , $ k = 0 , \cdots , (n-1)  $.  
We can check that these satisfy 
\bea\label{fundamental}  
&&   gh = h^q g \cr 
&&   g^n = h^n =1  
\eea 
These equations (\ref{fundamental}) are the fundamental equations 
used to describe the Dessins in \cite{jsw,jns}. 
We can also check $(gh)^n=1 $.

The check for $h^n$ is trivial. For
$g^n$ one writes out
\bea
g^{-n} ( nk +1 +i ) = n ( k+n  )_n + 1  + ( q^n i )_n
\eea
$(q^n i )_n = ( (q^n)_n (i )_n )_n   = i $.
Using Lemma 6 of \cite{jns} we have $(q^n)_n  =1 $.
We can also write out  the action of $(gh)$.
\bea
gh ( nk +1 +i ) = n (k-1)_n + 1 + ( {i+1 \over q} )_n
\eea
It is clear from this formula that
$n(k+1)_n   \le gh (x )  \le  n(k +2 )_n$
for $ nk \le x \le  n(k +1 ) $. We can also
check that $gh $ obeys $(gh)^n =1 $.  This proves that
$gh$ consists of $n$ cycles of length $n$.

From these fundamental permutations we get the Dessins \cite{jsw} as 
\bea 
&& \sigma_0 = g^u \cr 
&& \sigma_1 = (gh)^u 
\eea

As explained in the previous section \ref{sec:desstolists}, 
if we use the maximal number of colors, we can certainly 
reconstruct $\sigma_0 , \sigma_1$. The degree of the map 
is $n^2$, the cycle decompositions of $ \s_0 , \s_1$ are 
$[n^n]$, so we can extract $ n=p^e$.  
 To extract 
$f,u$ from the permutations, we calculate from the 
basic relations (\ref{fundamental})  that 
\bea
\sigma_1 \sigma_0^{-1}  = (  \sigma_0^{-1} \sigma_1 )^{q^u } 
\eea 
Hence the colored Dessins with maximal colors 
allow us to recover $   \sigma_0 , \sigma_1 $, 
and by comparing $ \sigma_1 \sigma_0^{-1}$ 
and $ \sigma_0^{-1} \sigma_1 $ we extract 
$q, u$, and in turn we extract $f$ from the definition $q = p^f +1 $.

This gives a proof-of-principle 
that the Galois invariants $p,e,f$ can be extracted 
using colorings of the Dessins. We have used the maximal 
number of colors which allow us to directly 
construct $ \sigma_0 , \sigma_1$. It would 
be interesting to find the minimal number of 
edge-colors which can reconstruct the data $p,e,f$.


\section{ Summary and Outlook }

Our motivation was to apply the logic  of the string theory of 2dYM
where  Hurwitz space was found to play a central role, to the
case of correlators  in hermitian matrix models.

\noindent
{\bf Summary of Key points} 
\begin{itemize}
\item[1.] By using diagrammatic tensor space calculations, 
we  showed that  correlators of arbitrary  multi-traces 
of the 1-matrix model at generic couplings
can be mapped to counting problems involving certain triples 
of permutations. This lead to  an interpretation in terms of Hurwitz numbers. 
 The matrix model has a dual string theory interpretation in terms 
of a target space $\mP^1$, 
which  localizes on holomorphic maps with three branch points.
Note that, in contrast to the traditional non-critical string
interpretation, this requires no double-scaling limit for a 
continuum interpretation.

\item[2.] We used results in the Matrix model literature to get
 explicit results on Hurwitz numbers with three branch points.
 We believe these results
on Hurwitz numbers are new and these calculations 
should admit many generalizations. 

\item[3.]  Exploiting one of Grothendieck's remarks \cite{grotesquisse},
 we highlighted the fact that
 the absolute Galois group organizes  the Feynman graphs of the 1-matrix model.
\item[4.] We observed that edge-colorings of Grothendieck
Dessins arise when we consider correlators of the multi-matrix models.
The counting of these edge-colored Dessins is related to 
counting of triples of permutations in $S_d$ 
with equivalences defined by subgroups. 
\item[5.] We argued that multi-matrix models at generic couplings 
are related to a continuum string
theory defined by  sheaves over the Hurwitz space of
branched covers with 3 branch points.
\item[6.] We used edge-colorings and related lists of multi-matrix operators
to define new  invariants of the Galois  action on Dessins.
\item[7.] We related known Galois invariants to edge-colorings and lists.

\end{itemize}

We will discuss extensions of this work, connections to recent literature
and intriguing puzzles under three main headings.  
\begin{itemize} 
\item{1.}  Hurwitz spaces, $ \cM_{g,n}$ and topological strings, 

\item{2.}  The target space of the string dual of Matrix models. 

\item{3.}  The absolute Galois group $ Gal ( \bmQ / \mQ ) $.  

\end{itemize}

{\bf Hurwitz spaces and topological strings }

 The computation (section \ref{newhurwitz}), using existing 
 Matrix model literature,  of explicit Hurwitz numbers
 with general ramification profiles over two branch points 
 but with simple ramification points over one branch point, 
 can clearly be generalized.  
The computation of explicit Hurwitz numbers  \cite{cretay} 
for a different case, namely where all branch points 
have simple ramification profile  of type $ [ 2 1^{d-2} ] $
 has been related recently to intersection theory over 
 $ \cM_{g,n} $ \cite{elsv}. An intersection theory approach to 
more general branching data for the case of genus zero world-sheet 
 is available \cite{shadrin}. Elaborating on the connection with the 
results of section \ref{newhurwitz} would be interesting. 

In the recent work \cite{Morozov:2009uy} 
exact answers for generating functions of 
correlators in the hermitian matrix correlators were constructed.
Given the results of section 2,  these
can be interpreted as generating functions 
for Hurwitz numbers. 

\noindent 
{\it  Different spacetimes from Refinements of Hurwitz counting }\\
$~~~ $ In this article we have focused on Matrix models which are 
zero dimensional quantum field theories. Similar 
calculations in higher dimensional conformal field theories, 
in the context of  the Zamolodchikov inner product on the space of 
all gauge invariant operators of interest in AdS/CFT, express correlators 
in terms of symmetric group data \cite{cjr,cr,dss,bhr, kr,  rob,bcd,withnick}. 
In \cite{cr} (for example section 5 and 8) 
there are computations showing how different refinements
of symmetric group  counting problems (hence Hurwitz space combinatorics)
are weighted with different space-time dependences
of correlators. This suggests that spacetime 
might emerge from refinements of Hurwitz counting problems.
Another  recent AdS/CFT development in connection with Hurwitz spaces is the 
work \cite{Pakman:2009mi}.  

{ \bf The target space of the dual string theory for the one-Matrix model   } 

Our work provides a continuum string interpretation
 of one-Matrix model correlators
at for arbitrary parameters of the potential $ V ( X ) $.  
The target space is  $\mP^1$.
The maps we encounter involve those where there are precisely 
three branch points on the target space. The ramification points 
in the inverse image of the branch points are determined by 
the parameters of the potential and the observable inserted
(see section \ref{universal}). 
Using the Belyi theorem, these curves and maps are 
defined over $ \bmQ$. This suggests that the target space is 
really  $\mP^1 ( \bmQ ) $.    

 The traditional  spacetime interpretation
 of the minimal strings is that we have one real dimension :
 which can be viewed as the Liouville direction, or as the
 eigenvalue direction of the Matrix model. 
A two dimensional target space, which is a Riemann 
surface semi-classically but becomes a  $\mP^1$
non-perturbatively was discussed in  \cite{mmss}. 
The  $\mP^1$ target space  is compatible 
 with our interpretation. While the observables we have considered 
 are powers of traces, the ones considered in \cite{mmss} 
 involve determinants. The geometrical structure we have 
 developed is Hurwitz spaces, whereas the geometries related to 
 the integrable structures there are infinite dimensional Grassmannians. 
 The Matrix theory transformation between the two types of  
 observables (see expressions in \ref{universal})
therefore seems to be capturing some interesting
 geometrical relations between Hurwitz spaces 
 for 3-branch point maps and Grassmannians.  
It would be very interesting to 
articulate that more precisely.

In the picture developed in this paper, 
the double-scaling limits can be viewed as arising 
by tuning the couplings of the different ramification points 
which are present in the exponential of the potential (\ref{expandexp}).  
So the string theory on $ \mP^1 (\bmQ )$ contains 
these different double scaled limits when the 
couplings of ramification points are tuned.   
This has some similarity to 
the  phase transition in 2dYM which was interpreted 
in terms of ``condensation of branch points''\cite{Taylor:1994zm}.

Our claim that we have the string theory
for the 1-matrix model at generic coupling is based
 on treating the partition function as
a perturbation of the Gaussian model.
In cases where  the perturbation series has a finite radius of
convergence,  one could argue that beyond the radius of
convergence, the description in terms of  string theory on $\mP^1$
 is not valid. This is reminiscent of the discussion of
phase transitions for 2dYM on the sphere \cite{kazak}.
Quite generally one can view the integral as a formal series
rather than a convergent series, so the 
picture of $\mP^1 ( \bmQ ) $ holds in this interpretation
(for relations between the convergent series and formal series 
 interpretations see  \cite{eynard}). 

String theory over fields other than $\mbC$ were discussed
in the past, especially  p-adic strings \cite{FO,VOL} 
and they have received renewed interest recently 
in the context of tachyon condensation \cite{GhoSen}.  
The observations in this paper 
are suggesting  that old-matrix models at generic couplings
admit such an interpretation as string theory over
 $ \bmQ$ with target space $ \mP^1 ( \bmQ )$. 
Is there a concrete construction of such a string ?
Since a lot of algebraic geometry such as that of Hurwitz spaces 
and $ \cM_{g,n}$ (see for example \cite{delignemumford}) 
is done over general algebraically closed fields,
 it is tempting to believe that the answer is yes. 
Is there an  explicit construction of a worldsheet string path integral 
over $\bmQ$ and the string field theory on $\mP^1 (  \bmQ ) $ ?

{ \bf The absolute Galois group }

The Galois group $ Gal ( \bmQ / \mQ ) $ 
is of fundamental interest in number theory. 
It has no known explicit description in terms of generators 
and relations for example. For mathematicians, the interest 
in the faithful action action of $ Gal ( \bmQ / \mQ ) $ on Dessins 
comes from the fact that it is a way to learn about the 
mysterious group itself. The coarser invariants 
of the Galois action, such as the conjugacy classes 
of the permutations in the description  in terms of 
triples of permutations, do not suffice to distinguish 
distinct Galois orbits. 

We have been lead, by considering Feynman diagrams of multi-matrix models, 
to study the Galois action on Dessins by doing what
infants would naturally do, namely color the edges. 
Coloring the edges captures combinatoric  information about the 
Dessins \footnote{Incidentally, another variation on this theme 
is that working with complex matrix models  amounts to 
coloring the edges as well as giving them an arrow.}. 
Some of this combinatoric information was related 
in section \ref{CEDGI} to known mathematical Galois invariants
 which contain information about how Dessins 
fit into Galois orbits. We also defined new invariants 
in terms of intersections and unions over Galois orbits 
of colored-edge Dessins. The union-invariants, with 
sufficiently many colors, are complete, in the sense 
that they can certainly contain enough information to 
characterize orbits. However, they contain 
in a sense too much information and computing
them would seem to  require scanning entire orbits. 
The intersection-invariant by contrast is local in that 
it is a property of the set of colorings of any single Dessin. 
An interesting problem is to find out if the intersection 
invariant is useful for distinguishing orbits that 
cannot be distinguished by other methods. 
More generally, we can ask whether there exists
any  complete set of  local  invariants,
which can be defined in terms of the colorings of a
given Dessin, and are complete in that they can always 
tell whether a pair of Dessins are in the same orbit or not.

\vskip1in

{ \Large
{ \centerline { \bf Acknowledgements }  } }

\vskip.2in
We thank  Amir Kashani Poor, Yusuke Kimura, Vishnu Jejjalla, David Madore, Rodolfo Russo,
David Turton  for useful discussions/correspondence.
SR is supported by an STFC grant ST/G000565/1.
RdMK is supported by the South African Research Chairs Initiative 
of the Department of Science and Technology and National Research Foundation.

\vskip.5in

\begin{appendix}

\section{ Glossary and key facts  } \label{glosskey}

\noindent
{\bf Absolute Galois group} \\ 
The Galois group of the closure $ \bmQ$ of  the rationals $ \mQ$. 
It is the subgroup of the automorphism group of 
$\bmQ$ which leaves  $\mQ$ fixed.

{\bf Automorphisms }\\ 
$~~$ $ Aut ( \s_0 , \s_1 ) $ is  the automorphism group of a set of triples
$\s_0,\s_1, (\s_0\s_1)^{-1} $
(defined in section (\ref{revrh})), $ Aut D $ the automorphism group of
a Dessin $D$, $ Aut f $ the automorphism group of a 
holomorphic map $f$ (defined in section \ref{OWP-3BPs}).
Under the correspondence between triples, Dessins and holomorphic maps 
with 3 branch points, these are isomorphic.

\noindent 
{\bf Belyi map } \\ 
$~~$ A map from $ \beta : \Sigma_h \rightarrow \mP^1 $ with 
three branch points at $0,1, \infty $ from a curve $ \Sigma_h$
(which is interpeted as the string worldsheet). The 
image  $\mP^1$ is the target space of the string theory.  
 
\noindent 
{\bf Belyi's theorem } \\ 
$~~$ All Belyi maps are  defined over $ \bmQ$. 
The underlying curve can be defined by algebraic 
equations involving algebraic number coefficients
and the map itself involves only such coefficients. 

\noindent
{\bf Bipartite graph } \\ 
$~~$ A graph with two types of vertices : Black and white. 
Black vertices are connected to white. The white vertices 
correspond to  ramification points above $1$. 
These graphs correspond to triples of permutations. 
The first permutation $ \sigma_0 $ described by 
the permutation of labeled edges around the black vertices 
as specified by the cyclic order which comes with each vertex. 
The second permutation $ \sigma_1$  describes permutations 
around the white vertices. See \cite{schneps,lanzvon,Joubert}.

\noindent
{\bf Branch point } \\
$~~$ A point on the target space, such that inverse image 
contains one or more points where the derivative vanishes. 
In terms of local coordinates $w$ on the target  
 there is at least one point in  the inverse image of $w=0$, 
with local coordinate $z$ on the worldsheet where the map is described 
by $w = z^i$  with  $i >1 $.{ \it CARE : } A branch point is a point 
on the target (unlike a ramification point).  

\noindent
{\bf Clean Belyi map } \\ 
$~~$ Belyi maps such that all ramification indices over the 
point $1$ are simple (look like $w = z^2$ in local coordinates).
 These maps are the sorts 
that arise in  the string description when we consider correlators in 
perturbed Gaussian matrix models. 

\noindent 
{\bf Clean Dessins } \\
$~~$ The white vertices each have two incident edges.  
Clean Dessins correspond to clean Belyi maps. 

\noindent
{\bf Clean Dessins $ \leftrightarrow $ Gaussian Matrix model  } \\
$~~$ The clean Dessins come from the Wick contractions in a Gaussian 
matrix model. Each trace ${\rm tr}X^k$ gives a black vertex with
$k$ edges. Each propagator is associated with a white vertex.
See section \ref{fromFeynDess}.

\noindent 
{\bf Colored-Edge-Dessins }  (CEDs) \\
$~~$ They are the Feynman graphs obtained in the 
multi-Matrix models, e.g a matrix model (Gaussian or perturbed Gaussian) 
with   different types of matrices $X , Y , Z \cdots $. 
They are related to triples of permutations in $S_d$ 
with equivalences defined by subgroups of 
$S_d$ (see definition in section \ref{MMCED}). 

\noindent
{\bf Complete bi-partite graph } \\ 
$~~$ Every white vertex is connected to every black vertex. 
Special case of interest to us in section \ref{sec:JSWinvts}  is $K_{n,n}$ 
where there is an equal number of black and white vertices. 
Many different Dessins can have the same underlying graph. 
Physicists are familiar with this fact from Feynman diagrams
in matrix models, where different connections between the same vertices
in a Feynman graph can change the $N$ dependence of the diagram.  

\noindent 
{\bf Correlators of Matrix operators } \\ 
$~~$ Correlators of the matrix operators are defined 
by insertion of the matrix operator in a 
hermitian matrix integral with a Gaussian action 
or perturbed Gaussian with a general potential 
parametrised by couplings $ g_3 , g_4 , \cdots $. 
After interating the Matrix we have the correlator.

\noindent
{\bf Delta function over symmetric group } \\ 
$~~$ For $ \sigma \in S_d$ we define $ \delta_{S_d}  ( \sigma ) =1 $ 
if $ \sigma $ is the identity permutation, and $ \delta_{S_d}  ( \sigma ) = 0  $
otherwise. By linearity this extends to a delta-function 
on the group algebra. This arises in counting branched covers 
(section (\ref{revrh}) ). 

\noindent 
{\bf Double Line diagrams, Ribbon graphs,  Grothendieck Dessins } \\ 
$~~$ Physicists usually use double line diagrams to describe 
the Feynman graphs of one-Matrix models. If we collapse 
the double lines to single lines, there is no loss of information 
as long as we keep  track of a cyclic order at the vertices, inherited from 
an oriented surface supporting the graph. These are ribbon diagrams 
or Grothendieck Dessins. In this paper, we mapped 
the computation of correlators in the Matrix model 
directly to counting of certain triples of permutations, 
which are well-known to be the combinatoric description of 
Dessins.

\noindent
{\bf From Belyi map to Dessin} \\ 
$~~$ The Dessin coresponding to a Belyi map 
is obtained as the inverse image of $[0, 1] $
under the Belyi map. Points over 1 are marked with white vertices
and points over 0 are marked with black vertices.  

\noindent 
{\bf Galois action on Dessins, or triples of permutations }\\
$~~$ Dessins or triples of permutations correspond to 
worldsheets and maps to the sphere $ \mP^1$ defined 
over $ \bmQ$. The Galois group acts on the elements of 
$ \bmQ$ and hence on Dessins or triples.

\noindent
{\bf Galois invariants of action on Dessins or triples  } \\
Examples of invariants are the conjugacy classes 
$ [\sigma_0 ] , [\sigma_1] , [\sigma_{\infty}]$, the 
automorphism group $ Aut ( D ) $. 
Finer invariants are discussed in section (\ref{CEDGI}).  

\noindent 
{\bf Galois closure of rationals   $ \bmQ  $ } \\ 
$~~$ The Galois closure of the rationals 
contains the solutions of all possible polynomial 
equations in a variable $x$ with coefficients 
which are rational, i.e in $ \mQ$. 

\noindent
{\bf Holomorphic map to $ \mP^1 $ } \\ 
$~~$ A map $ f : \Sigma_h \rightarrow \mP^1 $ 
from a Riemann surface of genus $h$ to the sphere $\mP^1$. 
Fixing a small disc  on the target space 
described by a local coordinate $ w$, the inverse image 
consists of a number of discs. Restricting to 
one disc, the map looks like $w = z^i$.
Summing over the indices $i$, we get the degree $d$ of the map. 

\noindent
{\bf Matrix operators } \\
$~~$ When computing a matrix integral over a matrix $X$, with a 
Gaussian or more general weight, observables of interest include
arbitrary moments built using traces and multi-traces such as 
$ ( tr (X))^3 $, $ tr X^3$. We call these observables matrix
operators. Multi-trace operators with $m$ copies of $X$ 
are related to conjugacy classes of $S_m$ (see section \ref{revtst}).
  
\noindent 
{\bf Multi-Matrix operators  } (MMOs)  \\ 
$~~$ They are traces of matrices involving multiple 
matrix types, e.g $  tr X^2trY^2 , trX^2Y^2 , tr XYXY $. 
Correlators of these multi-matrix multi-traces 
are the observables of interest in 
multi-matrix models. 

\noindent
{\bf Notation for conjugacy class of a permutation } \\ 
$~~$ If $ \s $ is a permutation,  $ [\s ]$ is the conjugacy class of the 
permutation, which is also the cycle structure of the permutation. 
 Conjugacy classed of $S_d$ are specified by partitions of $d$. 

\noindent 
{\bf Number field } \\ 
$~~$ Fields consists of a set of elements together with four operations,  
addition, subtraction, multiplication and division by nonzero elements. 
The rational numbers $\mQ$, together with their usual operations, form 
a number field.

\noindent
{\bf $\mP^1(\bmQ)$} \\
The projective line over the field $\bmQ$ is defined as the set 
of one-dimensional subspaces of the two-dimensional vector space  $\bmQ^2$.

\noindent
{\bf Ramification point } \\
$~~$ A point on the world-sheet where the derivative of 
the map vanishes. In the local description in 
terms of $ w = z^i $, we have $ i\ge 1 $. 
{\it CARE :}  A ramification point is a point on the worldsheet
(unlike a branch point). This usage of branch point and 
ramifcation point is conventional in mathematical  
literature on branched covers, but not by no means universal.  

\noindent
{\bf Ramification profiles } \\ 
$~~$ A set of positive integers describing the ramifications 
of all points in the inverse image of a branch point. 
The ramification profile of a branch point 
for a map of degree $d$ is a partition of $d$, 
which corresponds to conjugacy classes of $S_d$. 

\noindent
{\bf Riemann's existence theorem } \\ 
$~~$ Relates (equivalence classes of)
a sequence of permutations $ \sigma_1 , \sigma_2 , \cdots , \sigma_L$ 
obeying $ \sigma_1 \sigma_2 \cdots \sigma_L =1 $ 
to (equivalence classes of) holomorphic maps to 
$\mP^1$ target with $L$ branch points (see section \ref{revrh}). 

\noindent
{\bf Simple Hurwitz space} \\
$~~$ This is the space of holomorphic maps from worldsheet to 
target space, where the ramification profiles
over all branch points are of the form  $[2 1^{d-2}]$. 
This is the focus in  a number of discussion of 
 Hurwitz spaces in the context of topological string
theory and algebraic geometry. It is not the main subject of this paper. 

\vskip.1cm 
\noindent 
{\bf Triples of permutations } \\
$~~$ A set of three permutations 
$ \sigma_0 , \sigma_1 , \sigma_{\infty}  \in S_d $, 
such that $ \sigma_0 \sigma_1 \sigma_{\infty}  = 1 $. 
Equivalence of triples is defined in section \ref{revrh}. 
The computation of any observable in Gaussian 
or perturbed Gaussian matrix model can be mapped to 
a summation over equivalence classes of these triples
(see section \ref{OWP-3BPs}).

\noindent
{\bf Wick's theorem } \\ 
$~~$ A combinatoric rule which allows the computation 
of correlators in a Gaussian matrix model
and its perturbations by a general potential 
(see \ref{basiccorr}). It involves a sum over products of 
{ \it Wick contractions}. 

\end{appendix}

\end{document}